\documentclass[11pt,aps,amsmath,amssymb,nofootinbib,preprintnumbers, superscriptaddress,prd,notitlepage]{revtex4-1}

\usepackage{amsfonts}
\usepackage{amsmath}
\usepackage{amssymb}
\usepackage{epsfig}
\usepackage{latexsym}
\usepackage{paralist}
\usepackage{fancyhdr}
\usepackage{graphicx}
\usepackage{hyperref}
\usepackage{caption}
\usepackage{url}

\usepackage{color}
\usepackage{rotfloat}
\usepackage{float}
\usepackage{esint}
\usepackage{tikz}
\usepackage{etoolbox}

\numberwithin{equation}{section}

\usepackage[vcentermath]{youngtab}
\usepackage{young}
\usepackage{ytableau}
\usepackage{etex}
\usepackage{braket}
\usepackage{float}
\usepackage{slashed}
\usepackage{xcolor}
\usepackage{soul}

\usepackage{nicefrac}

\usepackage{tikzpagenodes}
\usepackage{lipsum}

\usepackage{multirow}
\usepackage{booktabs}

\usepackage{dcolumn} 
\usepackage{tabularx} 

\makeatletter
\@addtoreset{equation}{section}

\makeatother

\def\bea{\begin{eqnarray}} 
\def\eea{\end{eqnarray}}
\def\be{\begin{equation}} 
\def\ee{\end{equation}} 
\def\ba{\begin{array}}
\def\ea{\end{array}}

\def\be{\begin{equation}}
\def\ee{\end{equation}}
\def\bea{\begin{eqnarray}}
\def\eea{\end{eqnarray}}

\renewcommand{\thefootnote}{\fnsymbol{footnote}}

\let\oldtitle\title
\renewcommand{\title}[1]{\oldtitle{\color{blue}{#1}}}

\newlength{\myMheight}
\let\oldeqref\eqref
\let\oldcite\cite
\renewcommand{\eqref}[1]{{\color{blue}\oldeqref{#1}}}
\renewcommand{\cite}[1]{{\color{blue}\oldcite{#1}}}

\makeatletter
\let\reftagform@=\tagform@
\def\tagform@#1{\maketag@@@{\ignorespaces\textcolor{blue}{(\ignorespaces #1 \unskip\@@italiccorr \ignorespaces)\ignorespaces}}}
\makeatother

\makeatletter
\renewcommand{\p@subsection}{}
\renewcommand{\p@subsubsection}{}
\makeatother

\usepackage{mathpazo}
\usepackage{amsmath}

\begin{document}

\title{\textcolor{blue}{Platonic Field Theories}}

\author{R. Ben Al\`{i} Zinati}
\email{corresponding author: rbenaliz@sissa.it}
\affiliation{SISSA, International School for Advanced Studies \& INFN, via Bonomea 265, 34136 Trieste, Italy}

%
\author{A. Codello}
\affiliation{Department of Physics, Southern University of Science and Technology, Shenzhen 518055, China}
\affiliation{INFN - Sezione di Bologna, via Irnerio 46, 40126 Bologna, Italy}

\author{G. Gori}
\affiliation{Dipartimento di Fisica e Astronomia “Galileo Galilei”, Universit\`a di Padova, 35131 Padova, Italy}
\affiliation{CNR-IOM, via Bonomea 265, 34136 Trieste, Italy}

\begin{abstract}
\vspace{3mm}
We study renormalization group (RG)  fixed points of scalar field theories endowed with the discrete symmetry groups of regular polytopes. We employ the functional perturbative renormalization group (FPRG) approach and the $\epsilon$-expansion in $d=d_c-\epsilon$. The upper critical dimensions relevant to our analysis are $d_c = 6,4,\nicefrac{10}{3},3,\nicefrac{14}{5},\nicefrac{8}{3},\nicefrac{5}{2},\nicefrac{12}{5}$;
in order to get access to the corresponding RG beta functions, we derive general multicomponent beta functionals $\beta_V$ and $\beta_Z$ in the aforementioned upper critical dimensions, most of which are novel.
The field theories we analyze have $N=2$ (polygons), $N=3$ (Platonic solids) and $N=4$ (hyper-Platonic solids) field components.
The main results of this analysis include a new candidate universality class in three physical dimensions based on the symmetry group $\mathbb{D}_5$ of the Pentagon. Moreover we find new Icosahedron fixed points in $d<3$, the fixed points of the $24$-Cell, multi-critical $O(N)$ and $\phi^n$-Cubic universality classes.
\end{abstract}

\renewcommand{\thefootnote}{\arabic{footnote}}
\setcounter{footnote}{0}
\maketitle
\vspace{-10pt}
\section{Introduction}\label{sect:introduction}

The general problem of classifying universality classes of multicomponent scalar QFTs
is to date largely unsolved despite the centrality of the subject in modern days theoretical physics and the many decades passed since Wilson's original works \cite{WilsonKogut, WilsonFisher}.
In recent years the $\epsilon$-expansion has been reconsidered \cite{Osborn2018, Rychkov, Codello1} since it furnishes a simple method to 
approach the general classification of universality classes in arbitrary dimension, able to map uncharted territories in theory space, especially those pertaining to models having exotic or complex symmetry groups.
The analysis of single component scalar field theories with $\phi^k$ interactions teaches us
which are all  possible upper critical dimensions $d_c(k)$ around which the $\epsilon$-expansion can be performed.
Apart from the standard cases $\phi^3,\phi^4$ and $\phi^6$ corresponding, respectively, to integer $d_c = 6,4,3$ and which have been extensively studied \cite{Osborn1, brezin2, Zambelli, kompaniets2016, Adzhemyan:2019gvv, codello5, deAlcantaraPotts, deAlcantaraYangLee, Graceyphi3, Hager}, upper critical dimensions are generally rational and their universal leading order (LO) and next-to-leading order (NLO) contributions appear at loop orders higher than one; for this reason they have attracted attention only recently \cite{Osborn1, Codello1, Codello3, Gracey1}.

One of the main virtues of the functional reformulation of perturbative RG is the fact that multicomponent LO beta functionals, in any $d_c$, follow straightforwardly from their single component counterpart and thus no additional loop computations are needed to obtain the LO beta functions necessary for the fixed points (FPs) analysis. This important fact, for long time unnoticed, paves the way for the general analysis of multicomponent universality classes in dimension greater then two.
The typical approach to the  their classification
in the cases studied so far, i.e integer $d_c=6,4,3$, is to fix the number of components $N$ without assuming any symmetry for the models considered.
The analysis at fixed $N>1$ is a non-trivial algebraic problem in $\big ( {k+N-1 \atop k} \big)$ variables (number of marginal couplings)  and can be carried over in a fully analytical way only in the $N=2$ case (see \cite{Osborn2018} for the cases $d_c=6,4,3$ and \cite{Codello6} for the new case $d_c=\nicefrac{10}{3}$).
Higher number of components have been considered under the trace condition
in $d_c=4$ for  $N= 3, 4, 6$  \cite{Brezin, ZiaWallace, michel1, michel2, Hatch}, while the general problem in absence of this condition becomes rapidly algebraically intractable.
A complementary approach that will be pursued in this work is a ``symmetry perspective'' where one explores scalar theories characterised by a given family of symmetry groups $\mathcal{G}$s with the appropriate $N$-components representations and considering the upper critical dimensions implied by the functional form of the corresponding $\mathcal{G}$s-invariant Ginzburg-Landau (GL) Lagrangians.

Among the simplest families that exist for arbitrary $N$ and that have been the main object of study for decades, we recall the $O(N)$ symmetric theories in $d_c=4$, the Potts  $S_{N+1}$  families in $d_c=6$ and the Cubic$_N$ ones in $d_c=4$ (see \cite{Osborn2018} for a recent review and \cite{Pelissetto} for the state of the art). From a geometrical point of view, these symmetry groups correspond respectively to the $(N-1)$-sphere, the $N$-simplex and the $N$-cube. While the first is the simplest among continuous groups, the other two belong to the discrete group family of the regular polytopes and they are the only two which are present in any $N$-dimension\footnote{We refer to $N$-dimension as the dimension of the geometrical object considered  ruling the internal symmetry of our theory, which is not to be confused with the physical space dimension $d$.}.  
All the other regular polytopes can be constructed only in two (polygons), three (Platonic solids) and four (hyper-Platonic solids) $N$-dimensions.
In particular,  $N=2$ regular polytopes are the polygons and they are infinitely many.
In  $N=3$ we have only three cases up to duality: the Tetrahedron, the dual Octahedron/Cube pair and the dual Icosahedron/Dodecahedron  pair. 
Finally, in  $N=4$ there are four cases: the $5$-cell (hyper-Tetrahedron), the dual $8$-cell/$16$-cell pair (hyper-Cube/hyper-Octahedron), the $24$-Cell and  the dual $600$-cell/$120$-cell pair (hyper-Icosahedron/hyper-Dodecahedron).

In this paper we perform a systematic study of scalar field theories characterised by the symmetry groups of these geometrical objects.
Depending on the $N$-dimension considered, the related {\it Platonic Field Theory} (PFT) have order parameter with $N=2,3,4$ components and show up many possible upper critical dimensions; the ones we study are $d_c = 6,4,\nicefrac{10}{3},3,\nicefrac{14}{5},\nicefrac{8}{3},\nicefrac{5}{2},\nicefrac{12}{5}$.
We will look for fixed points of PFTs using the functional perturbative renormalization group (FPRG).
This can be achieved thanks to the aforementioned technical device that multicomponent beta functionals can 
be inferred from the knowledge of single component ones in a unique way at both LO and 
NLO in the `even' potential case and at LO in the `odd' potential case.

The paper is organised as follows. In Section \ref{sec:PFT} we define what we dub Platonic Field Theories (PFTs) introducing for each polytope (characterised by symmetry group $\mathcal{G}$) a method to construct basic $\mathcal{G}$-invariant polynomials which we use as building blocks to express the corresponding $\mathcal{G}$-invariant GL Lagrangian. We then determine the set of all possible upper critical dimensions $d_c$ the corresponding PFTs entail. 
In Section \ref{sec:MBF} we explain how to derive the beta functions for the marginal couplings generalising the single component beta functionals to their multicomponent version. 
The known cases of $d_c=6,4,\nicefrac{10}{3},3$ are reviewed and we give the new beta functionals for the cases $d_c=\nicefrac{8}{3},\nicefrac{14}{5}, \nicefrac{5}{2}, \nicefrac{12}{5}$ (the last two cases are given in Appendix \ref{Appendix:BetaFunctionals}).
In Section \ref{UC} we report a detailed analysis of all the fixed points and universality classes found (all the analytical details are contained in Appendix \ref{Appendix:AnalyticalDetails}). This section should be intended as a {\it guide map} to Table \ref{TFP} and Table \ref{TCE} which constitute the main results of this work and contain the relevant information regarding the critical behaviour of each polytope, namely for any admissible upper critical dimension, the corresponding fixed points and critical exponents. Concluding remarks and further perspectives are provided in Section \ref{sec:Conclusion}.

\section{Platonic Field Theories}\label{sec:PFT}

The  $N=2$ Platonic solids are nothing else than the regular polygons; a $n$-gonal regular polygon is represented by Schl\"afli symbol $\{ n \}$. 
$N=3$ Platonic solids are regular convex polyhedra: their faces are polygons $\{ p\}$, $q$ surrounding each vertex and they are denoted by Schl\"afli symbol $\{ p, q \}$.    The possible values of $p$ and $q$ can be enumerated and can have any other values than $\{ 3, 3 \}$, $\{ 3,4 \}$, $\{ 4,3 \}$, $\{ 3,5 \}$, $\{ 5,3 \}$ which identify the five Platonic solids in three dimensions.
Platonic solids in  $N=4$ (4-polytopes) are the analogs of the regular polyhedra in three dimensions and the regular polygons in two dimensions. The corresponding Schl\"afli symbol $\{ p, q, r \}$ identifies a solid with $\{ p \}$ faces and $\{ q , r \}$ vertex figures. The Schl\"afli's criterion \cite{coxeter} for the existence of a regular figure  corresponding to a symbol $\{ p, q, r \}$ selects the only 6 admissible 4-polytopes to be $\{ 3, 3, 3 \}$, $\{ 3, 3, 4 \}$, $\{ 4, 3, 3 \}$, $\{ 3, 4, 3 \}$, $\{ 3, 3, 5 \}$ and $\{ 5, 3, 3 \}$.
The symmetry groups $\mathcal{G}$ of the polytopes $\mathcal{P}$ considered are  listed in Table \ref{T-groups}. \\
\begin{table}
\begin{center}
\begin{tabular}{l c c c c c r}
		\hline\hline
		& & Polytope &  Schl\"afli	& & $\mathcal{G}$ & Molien Series $M(t)$\\
		\hline\hline 
		& & & &  \\
\multirow{1}{*}{$N=2$}		
		& &	$n$-Polygon	&	$\{n\}$   	& & 	$\mathbb{D}_n$ & $ [(1-t^2)(1-t^n)]^{-1}$ \\
		& & & &  \\ \hline \\
\multirow{5}{*}{$N=3$}
	& & Tetrahedron & $\{3,3\}$ & & $S_4$ & $[(1-t^2)(1-t^3)(1-t^4)]^{-1}$\\
	& \multirow{2}{*}{}  	& Octahedron & $\{3,4\}$	& &	$S_4\times\mathbb{Z}_2$  & \multirow{2}{*}{$[(1-t^2)(1-t^4)(1-t^6)]^{-1}$}\\
	& 				& Cube& $\{4,3\}$	& &	$S_4\times\mathbb{Z}_2$\\
	&  \multirow{2}{*}{} & Icosahedron & $\{3,5\}$	& &	 $A_5\times\mathbb{Z}_2$ & \multirow{2}{*}{$[(1-t^2)(1-t^6)(1-t^{10})]^{-1}$}\\
	& & Dodecahedron & $\{5,3\}$	& &	 $A_5\times\mathbb{Z}_2$\\
	& & & &  \\ \hline \\
\multirow{6}{*}{$N=4$}
		& &	5-cell & $\{3,3,3\}$ & & $S_5$ & $[(1-t^2)(1-t^3)(1-t^4)(1-t^5)]^{-1}$\\
		&  \multirow{2}{*}{} & $16$-cell	 &	$\{3,3,4\}$& &	 $(\mathbb{Z}_2)^4\rtimes S_4$ & \multirow{2}{*}{$[(1-t^2)(1-t^4)(1-t^6)(1-t^8)]^{-1}$}\\
		& & $8$-cell	 &	$\{4,3,3\}$	& &	 $(\mathbb{Z}_2)^4\rtimes S_4$\\
		& &	$24$-cell &	$\{3,4,3\}$	& & $F_4$ & $[(1-t^2)(1-t^6)(1-t^8)(1-t^{12})]^{-1}$\\
		&  \multirow{2}{*}{} & $120$-cell & 	$\{3,3,5\}$	& & $H_4$ &\multirow{2}{*}{$\quad\quad[(1-t^2)(1-t^{12})(1-t^{20})(1-t^{30})]^{-1}$} \\
		& & $600$-cell & 	$\{5,3,3\}$	& & $\,\,H_4$ \vspace{5pt}\\ 
		\hline\hline 
\end{tabular}
\end{center}
\caption{Polytopes symmetry groups $\mathcal{G}$ along with the corresponding Molien series.\\ The groups $F_4$ and $H_4$ are named according to the Coxeter notation.}
\label{T-groups}
\end{table}

In the RG  approach to critical phenomena, the critical behavior of PFTs can be described in terms of a $N$-component scalar field $\phi_i$ which carries an irreducible representation of a given polytope's symmetry group $\mathcal{G}$. Accordingly, the corresponding field theory will be described  by a GL action
\begin{equation}\label{GLA}
S = \int \mathrm{d}^d x \left\{ \frac{1}{2}\partial \phi_i \partial \phi_i+V(\phi_i) \right\}\,,
\end{equation}
where the GL potential $V(\phi_i)$  will be eventually expressed as a $\mathcal{G}$-invariant polynomial in the components $\phi_i$.
$\mathcal{G}$-invariant polynomials of degree $k$, namely $I^{(k)}(\phi_i)$,  can be constructed geometrically taking advantage of the strong symmetry of regular polytopes. 
To this purpose, let's consider the set of versors $\{ e^{\alpha} \}$ defining the $n$ vertices of a given polytope $\mathcal{P}$. 
In terms of these versors we construct the $k^{\text{th}}$ order invariant polynomial as\footnote{A regular polytope is easily seen to have a {\textit{centre}} from which all the vertices are at the same distance and therefore by construction it is  always true that $I^{(1)}=0$.}
\begin{equation}\label{eq:inv}
I^{(k\geq 2)}(\phi_i)=
\sum_{\alpha=1}^{n}e^{\alpha}_{a_1}\dots e^{\alpha}_{a_k}\, \phi_{a_1}\dots\phi_{a_k}\,,
\end{equation}
where summation over repeated indices is intended and we have chosen the versors to be normalized to 1. 
In general the explicit forms of the invariant polynomials $I^{(k)}$ depend on the choice of the (cartesian) coordinates which identify the vertices of $\mathcal{P}$, however, polynomials which are transformed into each other by a mere change of reference frame in the space of the $\phi_i$ components are physically equivalent and should not be distinguished.

\begin{table}[t]
\begin{ruledtabular}
\begin{tabular}{lccccccc}
$I^{(k)}$	& $\{3\}$	&	$\{4\}$	&	$\{5\}$	&	$\{6\}$	&	$\{7\}$	&	$\{8\}$\\
\hline\\
$I^{(2)}$	&$\rho=\eqref{rhotriangle}$ &	$\rho=\eqref{rhosquare}$	&	$\rho=\eqref{rhopentagon}$	&	$\rho=\eqref{rhohexagon}$	&	$\rho=\eqref{rhoheptagon}$	& 	$\rho=\eqref{rhooctagon}$	\\
$I^{(3)}$	&	$\tau=\eqref{tautriangle}$		&	$0$		&	$0$	&	$0$	&	$0$	&	$0$ \\
$I^{(4)}$	& & $\tau=\eqref{tausquare}$&	$\frac{3\rho^2}{10}$	&	$\frac{\rho^2}{4}$	&	$\frac{3\rho^2}{14}$	&	$\frac{3\rho^2}{16}$\\
$I^{(5)}$	& & &	$\tau=\eqref{taupentagon}$	&	$0$	&	$0$	&	$0$\\
$I^{(6)}$	& & &  &	$\tau=\eqref{tauhexagon}$	&	$\frac{5\rho^3}{98}$	&	$\frac{5\rho^3}{128}$\\
$I^{(7)}$	& & &  & &	$\tau=\eqref{tauheptagon}$ & 0\\
$I^{(8)}$	& & & & & &	$\tau=\eqref{tauoctagon}$	\vspace{5pt}\\
\end{tabular}
\end{ruledtabular}
\vspace{10pt}
\begin{ruledtabular}
\begin{tabular}{lccccccc}
$I^{(k)}$	& $\{3,3\}$ 	&	$\{3,4\}$	&	$\{3,5\}$	&	$\{3,3,3\}$	&	$\{3,3,4\}$	 &	$\{3,4,3\}$ \\
\hline\\
$I^{(2)}$	&$\rho=\eqref{rhotetra}$ &	$\rho=\eqref{rhoocta}$	&	$\rho=\eqref{rhoico}$	&	$\rho=\eqref{rho5cell}$	&	$\rho=\eqref{rho16}$	& 	$\rho=\eqref{rho24}$	\\
$I^{(3)}$	&	$\tau=\eqref{tautetra}$ & $0$ & $0$ & $\tau=\eqref{tau5cell}$	&	$0$	&	$0$ \\
$I^{(4)}$	& $\sigma=\eqref{sigmatetra}$ & $\tau=\eqref{tauocta}$&	$\frac{3\rho^2}{20}$	&	$\sigma=\eqref{sigma5cell}$	&	$\tau=\eqref{tau16}$	&	$\frac{\rho^2}{12}$\\
$I^{(5)}$	& & $0$ &	$0$	& $\omega=\eqref{omega5cell}$ &	0 &	$0$\\
$I^{(6)}$	& &  $\sigma=\eqref{sigmaocta}$&  $\tau=\eqref{tauico}$	& &	$\sigma=\eqref{sigma16}$	&	$\tau=\eqref{tau24}$\\
$I^{(7)}$	& & & $0$ & &	0 & 0\\
$I^{(8)}$	& & & $-\frac{7\rho^4}{960}+\frac{7\rho\tau}{15}$ & &$\omega=\eqref{omega16}$ &	$\sigma=\eqref{sigma24}$	\\
$I^{(9)}$	& & & $0$ & & & 0\\
$I^{(10)}$	& & & $\sigma=\eqref{sigmaico}$ &  & &	$\frac{7 \rho ^5}{41472}-\frac{7 \rho ^2 \tau }{144}+\frac{3 \rho  \sigma }{8}$	\\
$I^{(11)}$	& & & & & & 0 \\
$I^{(12)}$	& & &  &  & &	$\omega=\eqref{omega24}$
\end{tabular}
\caption{For each polytope $\mathcal{P}$, we give the basic $\mathcal{G}$-invariant polynomials  $I^{(k)}$, expressed in terms of the elements of the relative $B_\mathcal{P}$,  making reference to the corresponding equation in the main text. The Table makes clear the order at which the independent invariants appear. For any case related by duality, we give only the ones treated in the text and for simplicity we omit the $600$-cell case.}
\label{T-INV}
\end{ruledtabular}
\end{table}

Not all the invariants $I^{(k)}$ are independent, as can be inferred from Table \ref{T-INV}. For each polytope, we identify the basic $N$ independent ones by increasing the polynomial degree $k$. To this purpose it is useful to consider the Molien series which, for a given symmetry group $\mathcal{G}$, counts the number of homogeneous polynomials of a given degree $k$ that are invariants for $\mathcal{G}$ itself. It is defined as:
\begin{equation}
M(t)=\frac{1}{|\mathcal{G}|}\sum_{g\in \mathcal{G}}\frac{1}{\det[\mathbb{1}-t~\rho(g)]}\,,
\end{equation}
where $\rho$ is a linear representation of the group $\mathcal{G}$ on the underlying $N$ dimensional vector space. Once the series is expanded, the coefficient of the monomial $t^m$ gives the number of linearly independent homogeneous invariants of degree $m$; the Molien series furthermore suggests which is the polynomial degree of the basic $N$ independent invariant polynomials, as it can be understood cross-checking Tables \ref{T-groups} and \ref{T-INV}.
We always find only one quadratic independent invariant\footnote{A single quadratic invariant guarantees that the underlying fundamental representation of $O(N)$ remains irreducible under $\mathcal{G}$ and that we have only one phase transition.} which we call $\rho:=I^{(2)}$, while, independently of the order at which they first appear, we call $\tau$ the second and, when present, $\sigma$ and $\omega$ respectively the third and the fourth ones (see Table \ref{T-INV}). Let's call $B_{\mathcal{P}}$ the set given by the basic $N$ independent invariants of a given polytope $\mathcal{P}$. In terms of the elements of $B_{\mathcal{P}}$ we can consider $P^{(k)}(\rho, \tau,\sigma,\omega)$ as the most general homogeneous $\mathcal{G}$-invariant polynomial of degree $k$;
in general it can be expressed as
\begin{equation}
P^{(k)}(\rho, \tau,\sigma,\omega) = \sum_{\mu=1}^r g_\mu ~ M^{(k)}_{\mu} (\rho, \tau,\sigma,\omega)\,,
\end{equation}
where $M^{(k)}_{\mu} (\rho, \tau,\sigma,\omega)$ are monomials given by powers and products of elements of $B_\mathcal{P}$ such that their overall polynomial degree is $k$,  $g_\mu$ are some real coefficients  \cite{michel2} and the number $r$ of homogeneous polynomials of degree $k$ that are invariant under $\mathcal{G}_{\mathcal{P}}$, is given in terms of the Molien series as explained above.
In the framework of the $\epsilon$-expansion we are going to renormalize PFTs in $d=d_c-\epsilon$, where the upper critical dimension $d_c$ is uniquely determined by  the degree of the homogeneous polynomials $P^{(k)}$.  Indeed we can express the GL  $\mathcal{G}$-invariant potential $V(\phi_i)\equiv U (\rho, \tau,\sigma,\omega)$ simply as
\begin{equation}
U (\rho, \tau,\sigma,\omega) = \sum_{k=2} \frac{1}{k!}\, P^{(k)} (\rho, \tau,\sigma,\omega)\,,
\end{equation}
and we understand that the coefficients $g_{\mu}$ play the role of coupling constants.
By imposing the GL potential $U$ to be  marginal (remember that $\phi$ has dimensions $\frac{d-2}{2}$ as it can be gleaned out inspecting the kinetic part of the action \eqref{GLA}) we obtain the  upper critical dimensions as
\begin{equation}\label{ucds}
d_c(k)=\frac{2k}{k-2}\,.
\end{equation}
In this paper, for any polytope $\mathcal{P}$, we considered all the possible upper critical dimensions $d_c$ corresponding to the allowed $P^{(k\leq k_{\text{max}})}$, where $k_{\text{max}}$ is the degree of the highest order polynomial in $B_{\mathcal{P}}$. We exclude from the analysis those $d_c$ related to polynomials $P^{(k)}$ which are expressed as powers of $\rho$ only, since they will simply describe the corresponding $O(N)$ symmetric theory.

Let us make all this more concrete and give an example for the Square polygon $\{ 4 \}$.  First we construct the basic $\mathbb{D}_4$-symmetric invariant polynomials $I^{(k)}$. To this purpose, we fix the versors $\{e^{\alpha}\}$ choosing the four vertices of the Square to be the permutations of the coordinates $(\pm \nicefrac{1}{\sqrt2},\pm \nicefrac{1}{\sqrt2})$. We then proceed performing the sum in Eq. \eqref{eq:inv} which in this case extends up to $N=2$ and $n=4$. Starting from $k=2$ we find 
\begin{align}  
I^{(2)} &= 2 \left(\phi _1^2+\phi _2^2\right)\,,  \\
I^{(3)} &= 0 \,, \\
I^{(4)} &= \phi _1^4+6 \phi _2^2 \phi _1^2+\phi _2^4 \,,
\end{align}
and therefore the two elements of $B_{\{4\}}$ are $\rho_{\{4\}}\equiv I^{(2)}$ and $\tau_{\{4\}}\equiv I^{(4)}$.
Since the Square interaction term is represented by the invariant polynomial $\tau_{\{4\}}$ of degree $k=4$, the only interesting upper critical dimension in this case is $d_c=4$. The Molien Series for the Square group $\mathbb{D}_4$ is given by 
\begin{equation}
M(t) = [(1-t^2)(1-t^4)]^{-1} = 1 +t^2 +2 t^4 + 2 t^6 + O(t^8)\,,
\end{equation}
from which we understand that the $r=2$ monomials of degree $4$ are $M^{(4)}_{1}=\rho^2$ and $M^{(4)}_{2}=\tau$, so that the corresponding marginal potential $U(\rho,\tau)$ is given by
\begin{equation}\label{eq:squarepot}
U(\rho,\tau)=\frac{1}{4!}P_4(\rho,\tau)=\frac{1}{4!}\left(X \,\rho_{\{4\}}^2 + Y\, \tau_{\{4\}} \right)\,,
\end{equation}
where we named the coupling constants $g_1=X$ and $g_2=Y$.

As a further example, consider the case of the dual pair $\{ 3,4 \}$, $\{ 4,3 \}$ namely the Octahedron and the Cube.
We fix the versors $\{e^{\alpha}\}$ choosing the eight Cube vertices as the permutations of the coordinates $\sqrt{\nicefrac{4}{3}}(\pm 1,\pm 1, \pm 1)$ so that, once we perform the sum in Eq. \eqref{eq:inv} which now extends up to $N=3$ and $n=8$, we find that the three elements of $B_{\{4,3\}}$ in the Cube basis are
\begin{align}  
\rho_{\{4,3\}} &\equiv I^{(2)} = \frac{8}{3} \left(\phi _1^2+\phi _2^2+\phi _3^2\right)\,,\\
\tau_{\{4,3\}} &\equiv I^{(4)} = \frac{8}{9} \left(\phi _1^4+6 \left(\phi _2^2+\phi _3^2\right) \phi _1^2+\phi
   _2^4+\phi _3^4+6 \phi _2^2 \phi _3^2\right)\,,\\
\sigma_{\{4,3\}} &\equiv I^{(6)} =\frac{8}{27} \left(\phi _1^6+15 \left(\phi _2^2+\phi _3^2\right) \phi _1^4+15
   \left(\phi _2^4+6 \phi _3^2 \phi _2^2+\phi _3^4\right) \phi _1^2+\phi_2^6\right. \nonumber \\ &\left.
   \qquad\qquad+\phi _3^6+15 \phi _2^2 \phi _3^4+15 \phi _2^4 \phi _3^2\right)\,.
\end{align}
In the Octahedron basis the independent invariants are given in Appendix \ref{Appendix:AnalyticalDetails}. The duality between the two Platonic solids is expressed as a map between the invariants $(\rho,\tau,\sigma)$ in the two representations which, in the case of the Octahedron/Cube reads
\begin{eqnarray}\label{eq:octatocube}
\rho_{\{4,3\}} &=& \frac{4}{3}\,\rho_{\{3,4\}} \,,\nonumber \\
\tau_{\{4,3\}} &=& \frac{2}{3}\,\rho^2_{\{3,4\}} -\frac{8}{9}\,\tau^2_{\{3,4\}}\,, \nonumber \\
\sigma_{\{4,3\}} &=& \frac{5}{6}\,\rho^3_{\{3,4\}} -\frac{20}{9}\,\rho_{\{3,4\}}\tau_{\{3,4\}} + \frac{64}{27}\,\sigma_{\{3,4\}}\,.
\end{eqnarray}
The map between invariants translates in a smooth map between couplings and thus their RG properties are trivially the same.

Due to their interest in statistical physics \cite{Oshikawa, Delamotte1, AmitPeliti}, we notice as a final remark that $\mathbb{Z}_n$-symmetric models may be described in the long-distance limit in terms of a complex order parameter $(\phi,\bar{\phi})$ and mapped into a Lagrangian whose interaction term in general can be written as $(\lambda \phi^n+\bar{\lambda}\bar{\phi}^n)$.    
Imposing the reality of this interaction term amounts at enlarging the $\mathbb{Z}_n$ group to the corresponding dihedral one $\mathbb{D}_n$ and the $\mathbb{Z}_n$ invariants are nothing but the corresponding polygon ones. As an example consider the $\mathbb{Z}_5$ theory described by $(\lambda \phi^5+\bar{\lambda}\bar{\phi}^5)$; requiring $\lambda=\bar{\lambda}$ and changing representation to $\phi=\phi_1+i~\phi_2$, gives exactly the $\mathbb{D}_5$ Pentagon invariant considered in Eq. \eqref{taupentagon}.
\\

\section{Multicomponent beta functionals}\label{sec:MBF}

In order to study the RG flow of PFTs as presented in the previous section we use the perturbative formalism in its functional formulation  (FPRG) \cite{Osborn2018, Codello1}. 
In particular we use minimal subtraction scheme ($\overline{\text{MS}}$) in $d=d_c-\epsilon$ where, for each PFT, the upper critical dimensions $d_c$ are uniquely identified by Eq. \eqref{ucds} and specify the dimensions where to expect non-trivial universality classes. For each polytope the upper critical dimensions considered are listed in Table \ref{TFP}. 
The beta functions of the couplings appearing in the marginal potential $V(\phi)$ can be extracted from the beta functional $\beta_V$ while the flow of  $\beta_Z$ fixes the anomalous dimension $\eta$, where by  $Z(\phi)$ we denote a field-dependent wave-function (we refer to \cite{Codello1} for more details).

For even potentials, namely when $k=2m$ with integer $m>1$,  the upper critical dimensions $d_c$ in Eq. \eqref{ucds}  read $d_c=\frac{2m}{m-1}$ and the corresponding single component LO and NLO contributions are known in general  \cite{Osborn1}. LO beta functionals in the even case have been given recently 
for general $N$ in \cite{codello5}. 
While for $d_c=4$ and $d_c=3$ the NLO corrections are well known\footnote{In $d_c=4$ higher loop corrections are also known, but they are not universal and we do not consider them in the present paper.} \cite{Osborn2018}, there are no general expressions for the NLO multicomponent beta functionals for arbitrary $m$. But here is where the magic of the functional constraints comes to help. In fact, by analysing the form of the $N=1$ beta functionals given in \cite{Osborn1}, one realises that there is only one way to enhance them to the multicomponent case.

For example let's consider the $d_c=4$ case. The knowledge of the single component beta functionals  $\beta_V=\nicefrac{1}{2}(V^{(2)})^2-\nicefrac{1}{2}V^{(2)}(V^{(3)})^2$  and $\beta_Z=-\nicefrac{1}{6} (V^{(4)})^2$ leads directly to their multicomponent version since there is only way to "promote" the monomials to the $N>1$ case:   $(V^{(2)})^2 \to V_{a_1a_2}V_{a_1a_2}$ and  $V^{(2)}(V^{(3)})^2 \to V_{a_1a_2}V_{a_1a_3a_4}V_{a_2a_3a_4}$; similarly, taking care of the un-contracted indexes for $\beta_Z$, $(V^{(4)})^2 \to V_{a_1a_2a_3a_4}V_{a_1a_2a_3a_4}$. We finally obtain 
%
%
\begin{equation}\notag
d_c=4
\end{equation}
\begin{center}
\begin{tikzpicture}
\draw (0,0) circle (.5cm);
\filldraw [gray!50] (.5,0) circle (2pt);
\draw (.5,0) circle (2pt);
\filldraw [gray!50] (-.5,0) circle (2pt);
\draw (-.5,0) circle (2pt);
\draw (1.5,0) circle (.5cm);
\draw (1,0) to[out=0,in=180] (2,0);
\filldraw [gray!50] (1.5,.5) circle (2pt);
\draw(1.5,.5) circle (2pt);
\filldraw [gray!50] (1,0) circle (2pt);
\draw (1,0) circle (2pt);
\filldraw [gray!50] (2,0) circle (2pt);
\draw (2,0) circle (2pt);
\draw (3,0) circle (.5cm);
\draw (2.5,0) to [out=0,in=180] (3.5,0);
\filldraw [blue!50] (3.5,0) circle (2pt);
\draw (3.5,0) circle (2pt);
\filldraw [blue!50] (2.5,0) circle (2pt);
\draw(2.5,0) circle (2pt);
\end{tikzpicture}
\end{center}
\begin{equation}\label{BBetasd=4}
\begin{split}
\beta_{V}	&= \frac{1}{2}V_{a_1a_2}V_{a_1a_2}-\frac{1}{2}V_{a_1a_2}V_{a_1a_3a_4}V_{a_2a_3a_4}\\
(\beta_{Z})_{a_1a_2}  &= - \frac{1}{6}V_{a_1a_3a_4a_5}V_{a_2a_3a_4a_5}\,,
\end{split}
\end{equation}
\\
%
where we reported the corresponding perturbative diagrams using hereafter as a color code, grey for $\beta_V$'s and blue for $\beta_Z$'s. 
Similarly, in the $d_c=3$ case one can avoid performing a direct multicomponent computation simply generalizing $\beta_{V} = \frac{1}{3} (V^{(3)})^2 + \frac{1}{6} V^{(2)}(V^{(5)})^2 - \frac{4}{3} V^{(3)}V^{(4)}V^{(5)} - \frac{\pi^{2}}{12}(V^{(4)})^3$  as well as $\beta_{Z} = -\frac{1}{45} (V^{(6)})^2$ to the multicomponent case, namely

%
\begin{equation}\notag
d_c=3
\end{equation}
\begin{center}
\begin{tikzpicture}
\draw (0,0) circle (.5cm);
\draw (-.5,0) to [out=0,in=180] (.5,0);
\filldraw [gray!50] (.5,0) circle (2pt);
\draw (.5,0) circle (2pt);
\filldraw [gray!50] (-.5,0) circle (2pt);
\draw (-.5,0) circle (2pt);
\draw (1.5,0) circle (.5cm);
\draw (1,0) to[out=50,in=130] (2,0);
\draw (1,0) to[out=0,in=180] (2,0);
\draw (1,0) to[out=-50,in=-130] (2,0);
\filldraw [gray!50] (1.5,.5) circle (2pt);
\draw(1.5,.5) circle (2pt);
\filldraw [gray!50] (1,0) circle (2pt);
\draw (1,0) circle (2pt);
\filldraw [gray!50] (2,0) circle (2pt);
\draw (2,0) circle (2pt);
\draw (3,0) circle (.5cm);
\draw (2.531,-.171) to [out=-30,in=-150] (3.469,-.171);
\draw (2.531,-.171) to [out=30,in=150] (3.469,-.171);
\draw (3,.5) to[out=-90,in=150] (3.469,-.171);
\filldraw [gray!50] (3,.5) circle (2pt);
\draw (3,.5) circle (2pt);
\filldraw [gray!50] (3.469,-.171) circle (2pt);
\draw (3.469,-.171) circle (2pt);
\filldraw [gray!50] (2.531,-.171) circle (2pt);
\draw (2.531,-.171) circle (2pt);
\draw (4.5,0) circle (.5cm);
\draw (4.031,-.171) to [out=-30,in=-150] (4.969,-.171);
\draw (4.5,.5) to [out=-90,in=30] (4.031,-.171);
\draw (4.5,.5) to[out=-90,in=150] (4.969,-.171);
\filldraw [gray!50] (4.5,.5) circle (2pt);
\draw (4.5,.5) circle (2pt);
\filldraw [gray!50] (4.969,-.171) circle (2pt);
\draw (4.969,-.171) circle (2pt);
\filldraw [gray!50] (4.031,-.171) circle (2pt);
\draw (4.031,-.171) circle (2pt);
\draw (6,0) circle (.5cm);
\draw (5.5,0) to[out=50,in=130] (6.5,0);
\draw (5.5,0) to[out=0,in=180] (6.5,0);
\draw (5.5,0) to[out=-50,in=-130] (6.5,0);
\filldraw [blue!50] (5.5,0) circle (2pt);
\draw (5.5,0) circle (2pt);
\filldraw [blue!50] (6.5,0) circle (2pt);
\draw (6.5,0) circle (2pt);
\end{tikzpicture}
\end{center}
\begin{equation}\label{BBetasd=3}
\begin{split}
\beta_{V}	&= \frac{1}{3} V_{a_1a_2a_3}V_{a_1a_2a_3} + \frac{1}{6} V_{a_1a_2}V_{a_1a_3a_4a_5a_6}V_{a_2a_3a_4a_5a_6}\\
& -\frac{4}{3} V_{a_1a_2a_3}V_{a_3a_4a_5a_6}V_{a_1a_2a_4a_5a_6}-\frac{\pi^{2}}{12}V_{a_1a_2a_3a_4}V_{a_3a_4a_5a_6}V_{a_1a_2a_5a_6}\\
(\beta_{Z})_{a_1a_2}&=-\frac{1}{45} V_{a_1a_3a_4a_5a_6a_7}V_{a_2a_3a_4a_5a_6a_7}\,.
\end{split}
\end{equation}
%

\vspace{45pt}
We are now in the position to infer the beta functionals for the even potential's upper critical dimensions we are interested in, namely $d_c=\nicefrac{8}{3}, \nicefrac{5}{2}, \nicefrac{12}{5}$, generalising the single component ones given in \cite{Osborn1}. The result for $d_c=\nicefrac{8}{3}$ is given in Eq. \eqref{BBetasd=8/3}, while the cases $d_c=\nicefrac{5}{2}$ and $d_c=\nicefrac{12}{5}$ are given respectively in Eq. \eqref{Betasd=5/2} and Eq. \eqref{Betasd=12/5}. 
\!
%
%
\begin{figure}[h]
\begin{equation}\notag
d_c=\frac{8}{3}
\end{equation}
\begin{tikzpicture}
\draw (0,0) circle (.5cm);
\draw (-.5,0) to [out=50,in=130] (.5,0);
\draw (-.5,0) to[out=-50,in=-130] (.5,0);
\filldraw [gray!50] (-.5,0) circle (2pt);
\draw (-.5,0) circle (2pt);
\filldraw [gray!50] (.5,0) circle (2pt);
\draw (.5,0) circle (2pt);
\draw (1.5,0) circle (.5cm);
\draw (1,0) to [out=50,in=130] (2,0);
\draw (1,0) to [out=75,in=180] (1.5,.375);
\draw (1.5,.375) to [out=0,in=115] (2,0);
\draw (1,0) to [out=-75,in=180] (1.5,-.375);
\draw (1.5,-.375) to [out=0,in=-115] (2,0);
\draw (1,0) to [out=0,in=180] (2,0);
\draw (1,0) to[out=-50,in=-130] (2,0);
\filldraw [gray!50] (1.5,.5) circle (2pt);
\draw (1.5,.5) circle (2pt);
\filldraw [gray!50] (1,0) circle (2pt);
\draw (1,0) circle (2pt);
\filldraw [gray!50] (2,0) circle (2pt);
\draw (2,0) circle (2pt);
\draw (3,0) circle (.5cm);
\draw (2.531,-.171) to [out=-30,in=-150] (3.469,-.171);
\draw (2.531,-.171) to [out=30,in=150] (3.469,-.171);
\draw (3,.5) to[out=-45,in=115] (3.469,-.171);
\draw (2.531,-.171) to[out=65,in=180] (3,.15);
\draw (3,.15) to[out=0,in=115] (3.469,-.171);
\draw (2.531,-.171) to [out=0,in=180] (3.469,-.171);
\filldraw [gray!50] (3,.5) circle (2pt);
\draw (3,.5) circle (2pt);
\filldraw [gray!50] (3.469,-.171) circle (2pt);
\draw (3.469,-.171) circle (2pt);
\filldraw [gray!50] (2.531,-.171) circle (2pt);
\draw (2.531,-.171) circle (2pt);
\draw (4.5,0) circle (.5cm);
\draw (4.031,-.171) to [out=-30,in=-150] (4.969,-.171);
\draw (4.031,-.171) to [out=30,in=150] (4.969,-.171);
\draw (4.5,.5) to[out=-40,in=110] (4.969,-.171);
\draw (4.969,-.171) to[out=150,in=-90] (4.5,.5);
\draw (4.031,-.171) to [out=0,in=180] (4.969,-.171);
\filldraw [gray!50] (4.5,.5) circle (2pt);
\draw (4.5,.5) circle (2pt);
\filldraw [gray!50] (4.969,-.171) circle (2pt);
\draw (4.969,-.171) circle (2pt);
\filldraw [gray!50] (4.031,-.171) circle (2pt);
\draw (4.031,-.171) circle (2pt);
\draw (6,0) circle (.5cm);
\draw (5.531,-.171) to [out=-30,in=-150] (6.469,-.171);
\draw (6,.5) to[out=-40,in=110] (6.469,-.171);
\draw (6,.5) to[out=-90,in=30] (5.531,-.171);
\draw (6,.5) to[out=220,in=70] (5.531,-.171);
\draw (6.469,-.171) to[out=150,in=-90] (6,.5);
\filldraw [gray!50] (6,.5) circle (2pt);
\draw (6,.5) circle (2pt);
\filldraw [gray!50] (6.469,-.171) circle (2pt);
\draw (6.469,-.171) circle (2pt);
\filldraw [gray!50] (5.531,-.171) circle (2pt);
\draw (5.531,-.171) circle (2pt);
\draw (7.5,0) circle (.5cm);
\draw (7.031,-.171) to [out=-30,in=-150] (7.969,-.171);
\draw (7.031,-.171) to [out=30,in=150] (7.969,-.171);
\draw (7.5,.5) to[out=-90,in=30] (7.031,-.171);
\draw (7.969,-.171) to[out=150,in=-90] (7.5,.5);
\draw (7.031,-.171) to [out=0,in=180] (7.969,-.171);
\filldraw [gray!50] (7.5,.5) circle (2pt);
\draw (7.5,.5) circle (2pt);
\filldraw [gray!50] (7.969,-.171) circle (2pt);
\draw (7.969,-.171) circle (2pt);
\filldraw [gray!50] (7.031,-.171) circle (2pt);
\draw (7.031,-.171) circle (2pt);
\draw (9,0) circle (.5cm);
\draw (8.5,0) to [out=50,in=130] (9.5,0);
\draw (8.5,0) to [out=75,in=180] (9,.375);
\draw (9,.375) to [out=0,in=115] (9.5,0);
\draw (8.5,0) to [out=-75,in=180] (9,-.375);
\draw (9,-.375) to [out=0,in=-115] (9.5,0);
\draw (8.5,0) to [out=0,in=180] (9.5,0);
\draw (8.5,0) to[out=-50,in=-130] (9.5,0);
\filldraw [blue!50] (8.5,0) circle (2pt);
\draw (8.5,0) circle (2pt);
\filldraw [blue!50] (9.5,0) circle (2pt);
\draw (9.5,0) circle (2pt);
\end{tikzpicture}
\begin{equation}\label{BBetasd=8/3}
\begin{split}
\beta_{V}	&=\frac{1}{8} V_{a_1a_2a_3a_4} V_{a_1a_2a_3a_4} +\frac{1}{160} V_{a_1a_2} V_{a_1a_3a_4a_5a_6a_7a_8}V_{a_2a_3a_4a_5a_6a_7a_8}\\
& +\frac{9}{80} V_{a_1a_2a_3}V_{a_3a_4a_5a_6a_7a_8}V_{a_1a_2a_4a_5a_6a_7a_8}\\
&-\frac{3}{8} V_{a_1a_2a_3a_4}V_{a_2a_3a_4a_5a_6a_7a_8} V_{a_1a_5a_6a_7a_8}\\
&-\frac{\Gamma(\nicefrac{1}{3})^3}{24}  
    V_{a_1a_2a_3a_4a_5a_6}V_{a_1a_2a_3a_7a_8}V_{a_4a_5a_6a_7a_8}\\
&+\frac{3}{64} \left[\sqrt{3} \pi -3
   (2+\log 3)\right] V_{a_1a_2a_3a_4}V_{a_1a_2a_5a_6a_7a_8}V_{a_3a_4a_5a_6a_7a_8}\\
(\beta_{Z})_{a_1a_2}&=-\frac{1}{1120} V_{a_1a_3a_4a_5a_6a_7a_8a_9} V_{a_2a_3a_4a_5a_6a_7a_8a_9}\,.
\end{split}
\end{equation}
\end{figure}

\!\!
\noindent We underline two interesting aspects about these expressions: first, as can be noted from the diagrams above, they are of relatively high loop order since the LO contribution $\beta_V$ arises from a $(m-1)$-loop computation while the NLO functionals $\beta_V$  and $\beta_Z$ appear  at $2(m-1)$-loops; second all the coefficients reported are universal, i.e. independent of the specific RG scheme adopted. Even if it is not difficult to write down the beta functionals for general $m$ and $N$, their expressions become rapidly quite cumbersome and we won't report them here. In any case we have
checked that  the general LO contributions  agree with those recently derived by CFT methods in \cite{codello5} genersalising to the multicomponent case the results of \cite{Codello:2017qek}.

In the odd case where $k=2m+1$ with integer $m\geq 1$ and the upper critical dimensions read $d_c=2+\frac{4}{2m-1}$, we consider only the leading contributions for two reasons: first, as reported in \cite{Codello1} we have a general formula for the beta functionals only at LO; second, the enhancement from the single to the multicomponent case works only at LO for even theories, since the presence of higher powers of $V^{(2)}$ in the NLO beta functionals makes the $N=1$ case degenerate with respect to the multicomponent case.

The $d_c=6$ case is well known and the NLO contributions can be found in \cite{Osborn2018, deAlcantaraPotts}. We report here the LO contributions, which are those that can be inferred from the single component case
\begin{equation}\notag
d_c=6
\end{equation}
\begin{center}
\begin{tikzpicture}
\draw (1.5,0) circle (.5cm);
\filldraw [gray!50] (1.5,.5) circle (2pt);
\draw (1.5,.5) circle (2pt);
\filldraw [gray!50] (1.933,-.25) circle (2pt);
\draw (1.933,-.25) circle (2pt);
\filldraw [gray!50] (1.067,-.25) circle (2pt);
\draw (1.067,-.25) circle (2pt);
\draw (3,0) circle (.5cm);
\filldraw [blue!50] (3.5,0) circle (2pt);
\draw (3.5,0) circle (2pt);
\filldraw [blue!50] (2.5,0) circle (2pt);
\draw (2.5,0) circle (2pt);
\end{tikzpicture}
\end{center}
\begin{equation}\label{Betasd=6}
\begin{split}
\beta_{V}	&=-\frac{1}{6} V_{a_1a_2}V_{a_2a_3}V_{a_3a_1}\\
(\beta_{Z})_{a_1a_2}&= -\frac{1}{6}V_{a_1a_3a_4}V_{a_2a_3a_4} \,.
\end{split}
\end{equation}
%
The $d_c=\nicefrac{10}{3}$ single component case has been reported recently in \cite{Codello3}. The generalization to its multicomponent  version is straightforward and reads\footnote{We use a different normalization with respect to \cite{Codello3,Codello6}}
%
\begin{figure}[h]
\begin{equation}\notag
d_c=\frac{10}{3}
\end{equation}
\begin{tikzpicture}
\draw (0,0) circle (.5cm);
\draw (-.5,0) to [out=30,in=150] (.5,0);
\draw (-.5,0) to[out=-30,in=-150] (.5,0);
\filldraw [gray!50] (0,.5) circle (2pt);
\draw (0,.5) circle (2pt);
\filldraw [gray!50] (.5,0) circle (2pt);
\draw (.5,0) circle (2pt);
\filldraw [gray!50] (-.5,0) circle (2pt);
\draw (-.5,0) circle (2pt);
\draw (1.5,0) circle (.5cm);
\draw (1.5,.5) to[out=-90,in=30] (1.067,-.25);
\draw (1.5,.5) to[out=-90,in=150] (1.933,-.25);
\filldraw [gray!50] (1.5,.5) circle (2pt);
\draw(1.5,.5) circle (2pt);
\filldraw [gray!50] (1.933,-.25) circle (2pt);
\draw (1.933,-.25) circle (2pt);
\filldraw [gray!50] (1.067,-.25) circle (2pt);
\draw (1.067,-.25) circle (2pt);
\draw (3,0) circle (.5cm);
\draw (2.5,0) to [out=30,in=150] (3.5,0);
\draw (2.5,0) to[out=-30,in=-150] (3.5,0);
\filldraw [blue!50] (3.5,0) circle (2pt);
\draw (3.5,0) circle (2pt);
\filldraw [blue!50] (2.5,0) circle (2pt);
\draw(2.5,0) circle (2pt);
\end{tikzpicture}
\begin{equation}\label{Betasd=10/3}
\begin{split}
\beta_{V}&=\frac{3}{4} V_{a_1a_2}V_{a_1a_3a_4a_5}V_{a_2a_3a_4a_5}-\frac{27}{8} V_{a_1a_2a_3}V_{a_1a_4a_5}V_{a_2a_3a_4a_5} \\
(\beta_{Z})_{a_1a_2}  &=	-\frac{3}{40}V_{a_1a_3a_4a_5a_6}V_{a_2a_3a_4a_5a_6}\,.
\end{split}
\end{equation}
\end{figure}

\noindent Finally we analysed the $m=3$ case obtaining, as a new result, the beta functionals referring to the upper critical dimension $d_c=\nicefrac{14}{5}$; the result is as follows
%
\begin{figure}[H]
\begin{equation}\notag
d_c=\frac{14}{5}
\end{equation}
\begin{center}
\begin{tikzpicture}
\draw (0,0) circle (.5cm);
\draw (0.469,-.171) to[out=150,in=-90] (0,.5);
\draw (0,.5) to[out=-40,in=110] (0.469,-.171);
\draw (0,.5) to[out=-90,in=30] (-0.469,-.171);
\draw (0,.5) to[out=220,in=70] (-0.469,-.171);
\filldraw [gray!50] (0,.5) circle (2pt);
\draw (0,.5) circle (2pt);
\filldraw [gray!50] (0.469,-.171) circle (2pt);
\draw (0.469,-.171) circle (2pt);
\filldraw [gray!50] (-.469,-.171) circle (2pt);
\draw (-.469,-.171) circle (2pt);
\draw (1.5,0) circle (.5cm);
\draw (1,0) to [out=50,in=130] (2,0);
\draw (1,0) to [out=75,in=180] (1.5,.375);
\draw (1.5,.375) to [out=0,in=115] (2,0);
\draw (1,0) to [out=-75,in=180] (1.5,-.375);
\draw (1.5,-.375) to [out=0,in=-115] (2,0);
\draw (1,0) to[out=-50,in=-130] (2,0);
\filldraw [gray!50] (1.5,.5) circle (2pt);
\draw (1.5,.5) circle (2pt);
\filldraw [gray!50] (1,0) circle (2pt);
\draw (1,0) circle (2pt);
\filldraw [gray!50] (2,0) circle (2pt);
\draw (2,0) circle (2pt);
\draw (3,0) circle (.5cm);
\draw (2.531,-.171) to [out=-30,in=-150] (3.469,-.171);
\draw (3.469,-.171) to [out=150,in=-90] (3,.5);
\draw (2.531,-.171) to [out=30,in=150] (3.469,-.171);
\draw (2.531,-.171) to [out=0,in=180] (3.469,-.171);
\filldraw [gray!50] (3,.5) circle (2pt);
\draw (3,.5) circle (2pt);
\filldraw [gray!50] (3.469,-.171) circle (2pt);
\draw (3.469,-.171) circle (2pt);
\filldraw [gray!50] (2.531,-.171) circle (2pt);
\draw (2.531,-.171) circle (2pt);
\draw (4.5,0) circle (.5cm);
\draw (4.031,-.171) to [out=-30,in=-150] (4.969,-.171);
\draw (4.031,-.171) to [out=30,in=150] (4.969,-.171);
\draw (4.5,.5) to[out=-90,in=30] (4.031,-.171);
\draw (4.969,-.171) to[out=150,in=-90] (4.5,.5);
\filldraw [gray!50] (4.5,.5) circle (2pt);
\draw (4.5,.5) circle (2pt);
\filldraw [gray!50] (4.969,-.171) circle (2pt);
\draw (4.969,-.171) circle (2pt);
\filldraw [gray!50] (4.031,-.171) circle (2pt);
\draw (4.031,-.171) circle (2pt);
\draw (6,0) circle (.5cm);
\draw (5.5,0) to [out=50,in=130] (6.5,0);
\draw (5.5,0) to [out=75,in=180] (6,.375);
\draw (6,.375) to [out=0,in=115] (6.5,0);
\draw (5.5,0) to [out=-75,in=180] (6,-.375);
\draw (6,-.375) to [out=0,in=-115] (6.5,0);
\draw (5.5,0) to[out=-50,in=-130] (6.5,0);
\filldraw [blue!50] (5.5,0) circle (2pt);
\draw (5.5,0) circle (2pt);
\filldraw [blue!50] (6.5,0) circle (2pt);
\draw (6.5,0) circle (2pt);
\end{tikzpicture}
\end{center}
\begin{equation}\label{Betasd=14/5}
\begin{split}
\beta_{V}	&= -\frac{125}{72} V_{a_1a_2a_3a_4a_5a_6}V_{a_1a_2a_3a_7}V_{a_4a_5a_6a_7} + \frac{5}{144} V_{a_1a_2}V_{a_1a_3a_4a_5a_6a_7}V_{a_2a_3a_4a_5a_6a_7}\\
& +\frac{125}{144} V_{a_1a_2a_3}V_{a_1a_4a_5a_6a_7}V_{a_2a_3a_4a_5a_6a_7}\\
&-\frac{125(\sqrt{5}-1)\Gamma(\nicefrac{3}{10})\Gamma(\nicefrac{6}{5})}{96\, 2^{2/5}\sqrt{\pi}}V_{a_1a_2a_3a_4}V_{a_1a_2a_5a_6a_7}V_{a_3a_4a_5a_6a_7}\\
(\beta_{Z})_{a_1a_2}&=-\frac{5}{1008} V_{a_1a_3a_4a_5a_6a_7a_8}V_{a_2a_3a_4a_5a_6a_7a_8}\,.
\end{split}
\end{equation}
\end{figure}

\noindent As an example we show how to extract the beta functions in the case of the Square polygon $\{4\}$.
Since the upper critical dimension in this case is $d_c=4$, we then refer to the beta functional in Eq. \eqref{BBetasd=4} to obtain the couplings’ beta functions. To this purpose, consider the Square potential as defined in Eq. \eqref{eq:squarepot}  
 in terms of which we can define straightforwardly
\begin{equation}\label{eq:betaVsquare1}
\beta_V = \frac{1}{4!}\left( \beta_X~\rho^2_{\{4\}} + \beta_Y~\tau_{\{4\}}\right)\,.
\end{equation}
We then proceed computing the r.h.s of Eq. \eqref{BBetasd=4} which reads\footnote{We note here that functional derivatives are first taken w.r.t. the fields $\{\phi_i\}$ and then the result is re-expressed in the natural basis of the invariants $\{\rho,\tau\}$.}
\begin{align}\label{eq:betaVsquare2}
\frac{1}{2}V_{a_1a_2}V_{a_1a_2}-\frac{1}{2}V_{a_1a_2}V_{a_1a_3a_4}V_{a_2a_3a_4} = & -\frac{1}{54} \rho ^2 ~ X \left(256 X^2+6 X (24 Y-5)+9 Y (2 Y-1)\right)\notag \\
& -\frac{1}{36} \tau ~ Y
   (8 X+3 Y) (32 X+12 Y-3)\,.
\end{align}
One then inserts \eqref{eq:betaVsquare1} and \eqref{eq:betaVsquare2}, respectively, on the l.h.s. and r.h.s. of  Eq. \eqref{BBetasd=4} and equates equal powers of the invariants on both sides to read off the corresponding dimension-full beta functions. Switching to dimensionless variables is straightforward\footnote{With abuse of notation we use the same symbols for dimensionless and dimensional couplings.} and the resulting system of beta functions is given in Eqs. \eqref{betaXsquare} and \eqref{betaYsquare}.

\section{Universality Classes}\label{UC}
The result of our analysis is reported in Table \ref{TFP}, which together with Table \ref{TCE}, are the main results of this work. This Section should be intended as the guide to these two Tables which the reader should have at hand.  
Table \ref{TFP} is basically composed of three columns: the first lists the polytopes; 
the second one reports the upper critical dimensions examined, which we remember are those where the relative PFT homogeneous invariant polynomials $P^{(k)}$ (interactions) are marginal (see Section \ref{sec:PFT}); the third one lists all real FPs found, i.e. all the real zeros of the corresponding  system of beta functions, whose solutions are labelled with the name of the {\tt{universality class}}\footnote{We use {\tt{typewriter}} font to denote universality classes.}  to which they correspond. Table \ref{TCE} instead reports the critical exponents $\eta$ and $\nu$ for all those universality classes for which we were able to compute both of them.

We start our analysis considering the polygons, namely the $N=2$ case.
Since there are an infinite number of polygons, we limited our analysis up to the Octagon, which is enough to show the general critical pattern emerging from the two families of {\textit{even}} and {\textit{odd}} $n$-gons. 
The Triangle in $d_c=6$ is the well known {\tt{Potts}$_3$} \cite{GolnerPotts, AmitPotts1, ZiaPotts, AmitPotts2, deAlcantaraPotts} which has a real FP
(but note the unusual fact: $\nu<\nu_{\text{MF}}=\nicefrac{1}{2}$).
It is well known that {\tt{Potts}$_3$} is not present in $d=3$ \cite{Nienhuis}, and this is an indication that even near $d=6$  it doesn't have a clear status (one can construct an argument using the NLO beta functions to claim the same \cite{AmitPotts1, AmitPotts2}).
The Square FPs in $d_c=4$ are the {\tt{O(2)}} and two copies of {\tt{Ising}}. Particular to the $N=2$ case is a mapping in terms of which it is true that {\tt{Cubic}$_2$=\tt{Ising}} \cite{Osborn2018, Pelissetto} and therefore the cubic FP is not present in this case. Cubic FPs emerge instead in the $N=3$ and $N=4$ cases as we shall see below.
The first surprise among polygons is the {\tt{Pentagon}} universality class. The upper critical dimension in this case is $d_c=\nicefrac{10}{3}$ and therefore it is a candidate to give a non-trivial critical behavior in three dimensions. The corresponding critical exponents are reported in Table \ref{TCE}. It is reassuring to see that $\nu > \nu_{\rm MF}$ contrary to what found in the single field case \cite{Codello3} for this upper critical dimension. Note also that the anomalous dimension is quite large in $d=3$ where it assumes the value $\eta=\nicefrac{1}{5}$;
it is natural therefore to consider this universality class in three dimensions where the $\epsilon$-expansion may have well behaved convergence properties since we just have to set $\epsilon = \nicefrac{1}{3}$.
The next polygon is the Hexagon which is analysed in $d_c=3$. In this case only a FP which identifies the tri-critical version of the {\tt{O(2)}}, namely the {\tt{Tri-O(2)}}, is present. The corresponding anomalous dimension is $\eta=\frac{1}{392}\epsilon^2$. 
The {\tt{Heptagon}} case in $d_c=\nicefrac{14}{5}$ shows a behaviour analogous to the {\tt{Pentagon}}, namely there is real FP representative of this universality class with 'well behaved' critical exponents given in Table \ref{TCE}. Even though this  universality class is new, it is less interesting w.r.t. the {\tt{Pentagon}} one since it does not exist in three dimensions and possibly exists only in $d=2$. 
Finally we analysed the Octagon in $d_c=\nicefrac{8}{3}$ which exhibits a critical behaviour analogous to the Hexagon case. In particular we find only the tetra-critical version of the {\tt{O(2)}} FP, namely only the {\tt{Tetra-O(2)}}, with anomalous dimension given by $\eta=\frac{9}{59858}\epsilon ^2$. We expect the \textit{even} family of $n$-gons with $n>8$ to reproduce the series of multi-critical {\tt{O(2)}} FPs\footnote{In particular, an even $n$-gon is characterised by the $\nicefrac{n}{2}$-th  multi-critical {\tt{O(2)}} FP.}.
Even though, within the formalism presented in Section \ref{sec:MBF}  and in Appendix \ref{Appendix:BetaFunctionals}, we could consider polygons $\{ n \}$ with $n > 8$ being {\textit{even}}  or {\textit{odd}} generalising to the multicomponent case the beta functionals of \cite{Osborn2018, Codello1}, we will not pursue this analysis here. It is anyway of interest to understand if the appearing of a non-trivial FP as for the {\tt{Pentagon}} and the {\tt{Heptagon}} is a general feature of all the {\textit{odd}} $\{ n \}$ theories or if there is a critical number of edges after which the fluctuations drive the FP to the corresponding {\tt{O(2)}} universality class.

We now move to the $N=3$ case, where we encounter the famous five Platonic solids.
We first analyzed the Tetrahedron which belongs to the family of simplexes; in this case the possible upper critical dimensions are $d_c=6$ and $d_c=4$. The first gives rise to no real FP, mirroring the fact that no real {\tt{Potts}$_4$} FP is known in $d\geq3$  \cite{Nienhuis}; in $d_c=4$, due to the fact that for $N=3$ the tetrahedral group is isomorphic to the cubic one ($\mathcal{G}=S_4\times\mathbb{Z}_2$, see Table \ref{T-groups}), the tetrahedral FPs coincide with the cubic ones \cite{ZiaPotts, Osborn2018}. The three universality classes that emerge are therefore $3\times${\tt{Ising}}, {\tt{O(3)}} and {\tt{Cubic}$_3$}, a case that has been extensively studied \cite{Aharony1, Aharony2, WallaceCubic, Pelissetto, Calabrese}. 
We considered the Cube/Octahedron pair in the Octahedron basis where the invariant polynomials assume a simpler form, see Appendix \ref{Appendix:AnalyticalDetails}. As explained above, due to the group isomorphism between the Tetrahedron and the Cube, the universal content in $d_c=4$ coincides. The second allowed upper critical dimension is $d_c=3$ where we find the tri-critical version of the previous FPs. In particular the $\phi^6$-{\tt{Cubic}}$_3$  FP is new and should be intended as a $\phi^6$-theory with cubic symmetry\footnote{In order to determine the exact degree of multi-criticality one has to analyse the corresponding stability matrix.}. Its critical exponents are given in Table \ref{TCE}. 
While it is clear \cite{Stergiou,Osborn2018, Delamotte1}  that no icosahedral FPs can be found in $d=4-\epsilon$ since the first invariant polynomial is of degree $6$, our analysis revealed that as we study the icosahedral theory in $d_c=\nicefrac{5}{2}$ by means of the marginal potential in Eq. \eqref{icopot5/2},  there emerge two icosahedral FPs for which we were able to compute the anomalous dimensions
%
\begin{eqnarray} \label{eq:icoFP}
\eta_1 &=& 7.95024 \times 10^{-6} \,\epsilon^2 \,,\\
\eta_2 &=& 8.93795 \times 10^{-6} \,\epsilon^2 \,.
\end{eqnarray}
%
Apart from the new icosahedral FPs, we find the {\tt{Tri-O(3)}} FP in $d_c=3$, the {\tt{Tetra-O(3)}} FP in $d_c=\nicefrac{8}{3}$ and finally the {\tt{Penta-O(3)}} FP in $d_c=\nicefrac{5}{2}$, the last two being new to our knowledge. The critical exponents are reported in Table \ref{TCE} while for the {\tt{Penta-O(3)}} FP we computed only the anomalous dimension $\eta=\frac{231}{25694761}\epsilon^2$, due to the high complexity of the NLO terms. We have analyzed the Icosahedron/Dodecahedron pair in the icosahedral basis; details on the invariant polynomials and on the duality map can be found in Appendix \ref{Appendix:AnalyticalDetails}.

Finally we considered the $4$-polytopes, namely the $N=4$ hyper-Platonic solids.
The PFT associated to the 5-cell (hyper-Simplex) entails upper critical dimensions $d_c=6,4,\nicefrac{10}{3}$. As expected, to the 5-cell in $d_c=6$ corresponds no real FP \cite{Nienhuis}.
In $d_c=4$ instead, apart from the {\tt{O(4)}} symmetric FP, the restricted Potts case gives rise to a  {\tt{Quartic-Potts}$_5$} FP  \cite{Osborn2018}, while the new information is that there is no real FP in $d_c=\nicefrac{10}{3}$.    
We explored 8-cell/16-cell symmetry in the basis of the 16-cell (hyper-Octahedron), where invariant polynomials are much simpler; details on the duality map are given in Appendix \ref{Appendix:AnalyticalDetails}. The analogy is perfect with the $N=3$ cubic case apart from the fact that no {\tt{Cubic$_4$}} FP is present in $d_c=4$ but only $4\times${\tt{Ising}} and {\tt{O(4)}} universality classes emerge \cite{Osborn2018}. In $d_c=3$  the FPs correspond to the $\phi^6$-{\tt{Cubic}}$_4$ and to the tri-critical version of $4\times${\tt{Ising}} and {\tt{O(4)}} while in $d_c=\nicefrac{8}{3}$ they simply are the $\phi^8$-{\tt{Cubic}}$_4$  and the tetra-critical version of $4\times${\tt{Ising}} and  {\tt{O(4)}}. 
All the new critical exponents are reported in the Table \ref{TCE}.
The 24-cell symmetry is peculiar to the $N=4$ case. In $d_c=3$ and $d_c=\nicefrac{8}{3}$ we respectively find only the {\tt{Tri}-O(4)} and the {\tt{Tetra}-O(4)} FPs.
Beside the {\tt{Penta}-O(4)} with anomalous dimensions given by $\eta=\frac{1}{83544}\epsilon^2$,  in $d_c=\nicefrac{5}{2}$ we find two 24-cell FPs characterized by the same anomalous dimension\footnote{These two FPs can be related by an $O(4)$ field redefinition.}
%
\begin{equation} 
\eta_{1,2} =1.15365\cdot10^{-5}\,.
\end{equation}
%
In $d_c=\nicefrac{12}{5}$ instead, there emerge two distinct 24-cell FPs with anomalous dimensions given by
%
\begin{align*} 
\eta_1 &=3.77524\cdot10^{-7}\,,\\
\eta_2 &=0.00171903 \,,
\end{align*}
%
along with the {\tt{Hexa}-O(4)} FP, whose anomalous dimensions reads $\eta=\frac{25}{53014528}\epsilon^2$.
Any of these universality classes can be present only in two dimensions. 
The analysis of the $600$-cell/$120$-cell symmetry starts to be very complicated even if straightforward. We considered this dual pair of polytopes at the upper critical dimensions  $d_c=\nicefrac{12}{5}, \nicefrac{20}{9}, \nicefrac{15}{7}$, but for simplicity we omit to report the corresponding invariants and beta functions though easily accessible along the line of reasoning of Sections \ref{sec:PFT} and \ref{sec:MBF}.
The analysis of the FPs for this dual pair revealed no other real FP except for the multi-critical {\tt{O(4)}} FPs (see Table \ref{TFP})\footnote{In Table \ref{TFP} we called {\tt{Triaconta}-O(4)} the multi-critical {\tt{O(4)}} FP associated to a $\phi^{30}$ theory.}. This result is somehow expected since, due to the high number of points on the unit 4-sphere it can be considered very close to the {\tt{O(4)}} model. We notice here that we could have analyzed the $600$-cell/$120$ pair in all the intermediate accessible upper critical dimensions, but since the analysis at the highest polynomial of degree $k_{\text{max}}=30$ revealed no $600$-cell FP, we expect the aforementioned $d_c$ to correspond only to the $O(4)$  multi-critical FPs.

It is natural to consider extensions of the present analysis based on the regular $N$-polytopes for general $N\geq 5$. However, we have that apart from $N$-simplexes (hyper-Tetrahedra) studied in \cite{Codello6}, just hyper-Cubes (hyper-Octahedra) are present and both their critical content is, on the other hand, already known.

\newpage

\begin{table}[H]
\begin{ruledtabular}
\begin{tabular}{lccccr}
		& & \!\!\!\!\! \!\!\!\!\! \!\!\!\!\! \!\!\!\!\!\!\!\!\!\! \!\!\!\!\!\!\!\!\!\!Polytope &  & $d_c$	& 	Fixed Points \\ \hline \\
\multirow{9}{*}{$N=2$}		
		& \parbox[c]{1em}{\includegraphics[scale=.1]{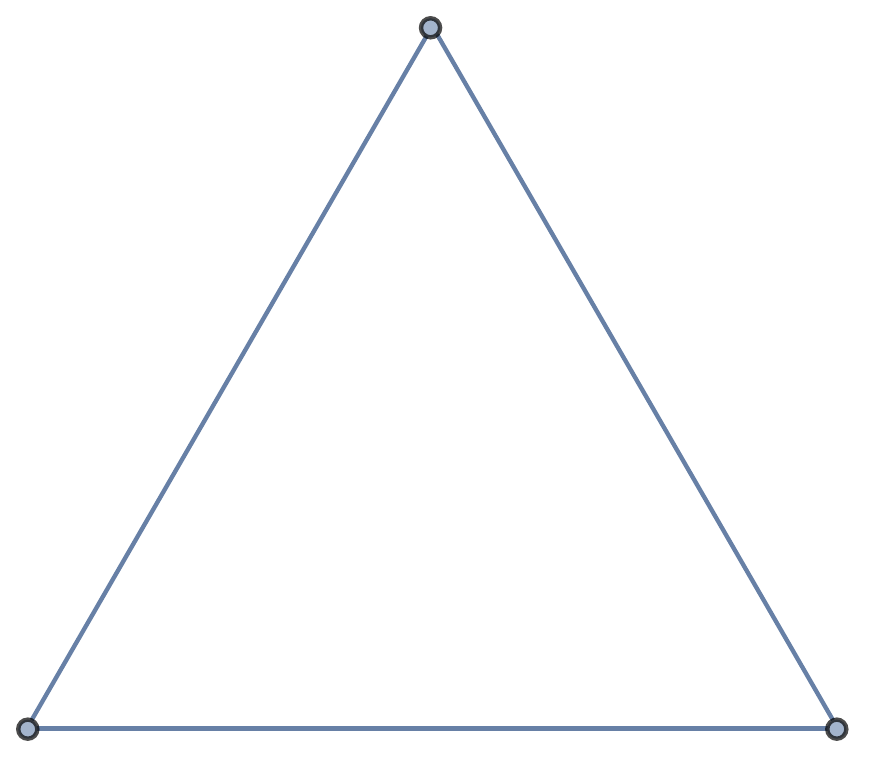}} &	Triangle	&	    	& 	$6$ 	& {\tt{Potts}$_3$} \\
		& \parbox[c]{1em}{\includegraphics[scale=.1]{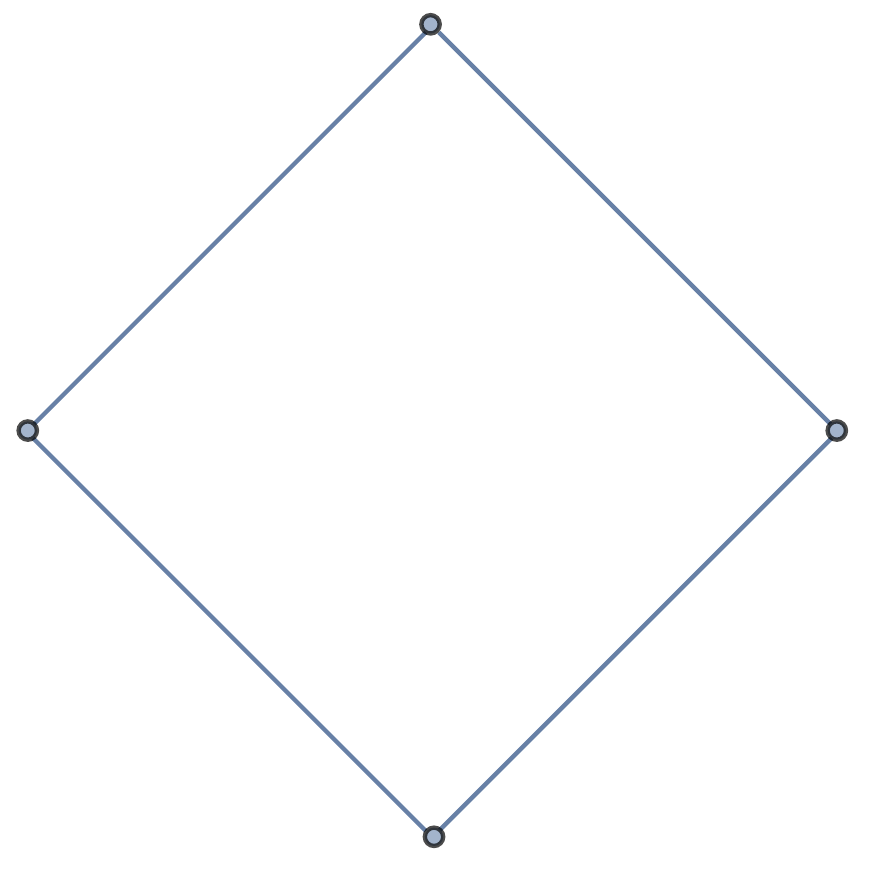}} &	Square	&      	 		&	$4$   	& {\tt{$2\times$Ising}},\, {\tt{O(2)}}  \\
            	& \parbox[c]{1em}{\includegraphics[scale=.1]{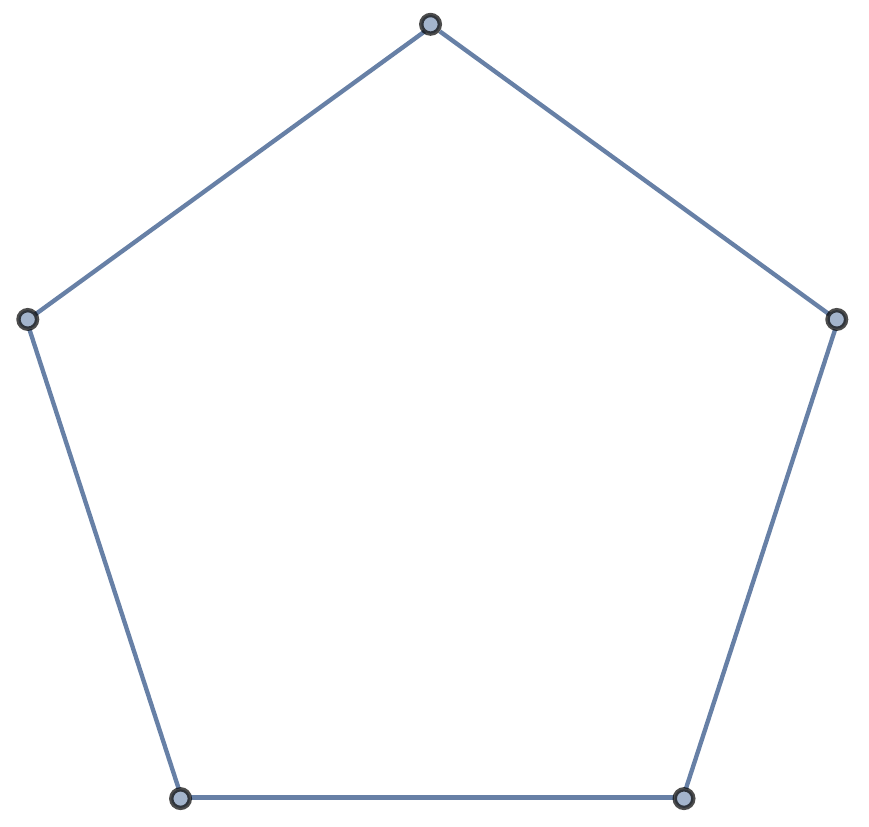}} &	Pentagon	&   	 		&	$\nicefrac{10}{3}$  	& {\color{black}{\tt{Pentagon}}} \\
               	& \parbox[c]{1em}{\includegraphics[scale=.1]{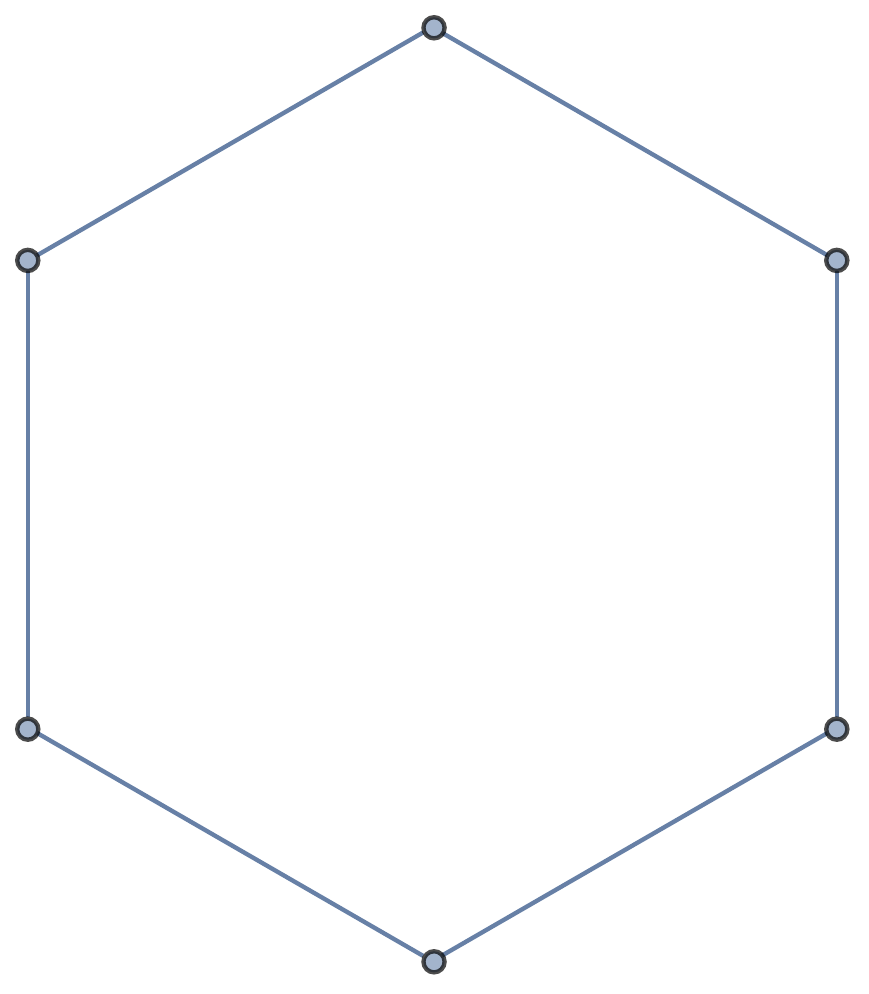}} &	Hexagon	&     	 		&	$3$   	& {\tt{Tri-O(2)}} \\
         	& \parbox[c]{1em}{\includegraphics[scale=.1]{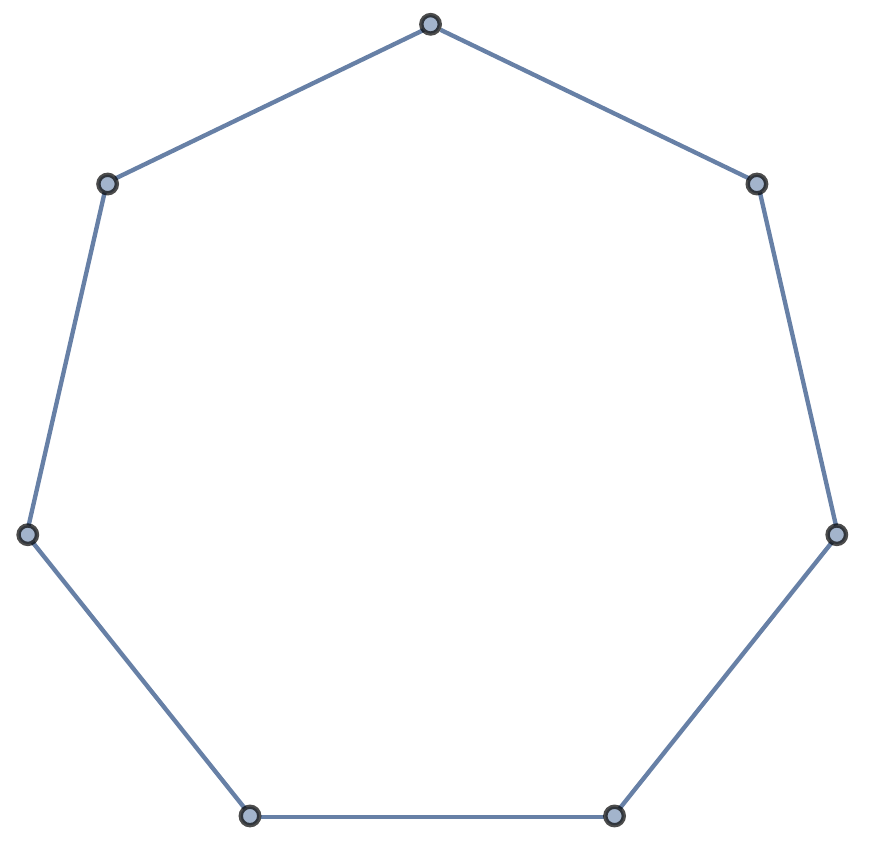}} &	Heptagon	& 	 		&	$\nicefrac{14}{5}$  	& {\color{black}{\tt{Heptagon}}} \\
	         & \parbox[c]{1em}{\includegraphics[scale=.125]{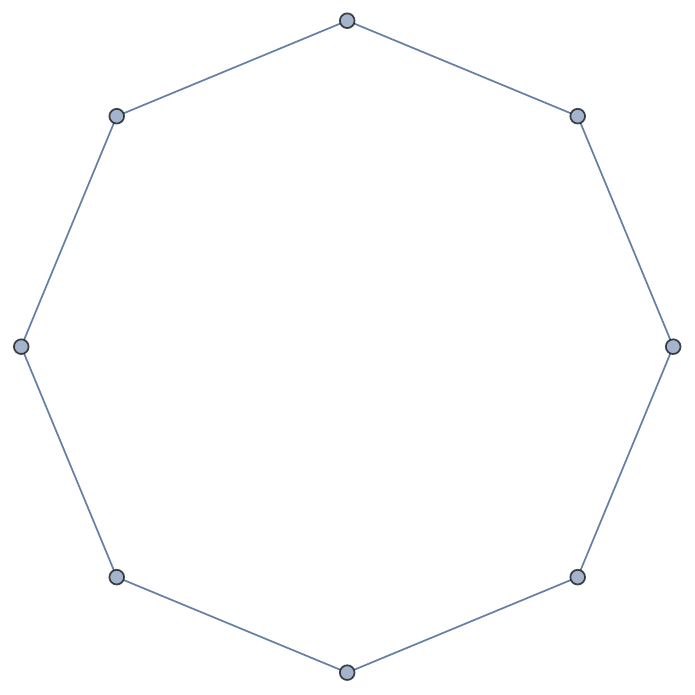}} &	Octagon	& 	 		&	$\nicefrac{8}{3}$  	& {\tt{Tetra-O(2)}} \\
	& $\quad\;\,\vdots$ &	$\vdots$ 	& 			&	  $\vdots$	& $\vdots\qquad$  \\
\\ \hline \\
\multirow{8}{*}{$N=3$}
	& \multirow{2}{*}{\parbox[c]{1em}{\includegraphics[scale=.1]{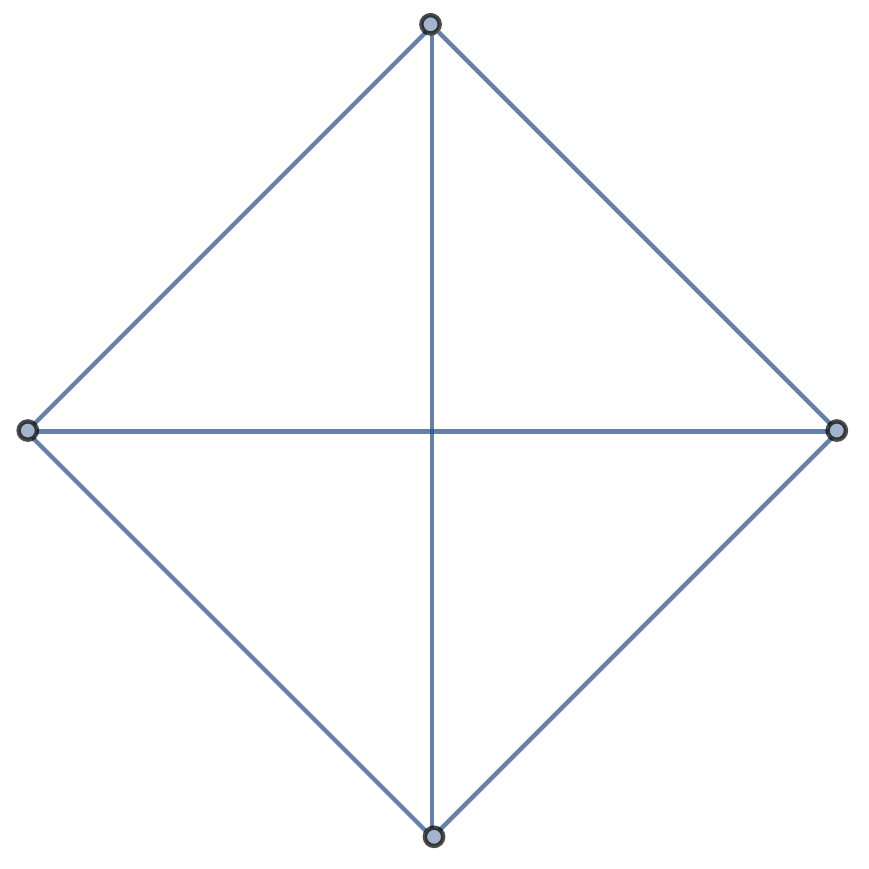}}}	 &	\multirow{2}{*}{Tetrahedron}	 & & $6$ &  {\tt{No real FP}}\\
		& & & & $4$ & {\tt{$3\times$Ising}},\, {\tt{O(3)}},\,  {\tt{Cubic}$_3$} \\
		& \\
		& \multirow{2}{*}{\parbox[c]{1em}{\includegraphics[scale=.1]{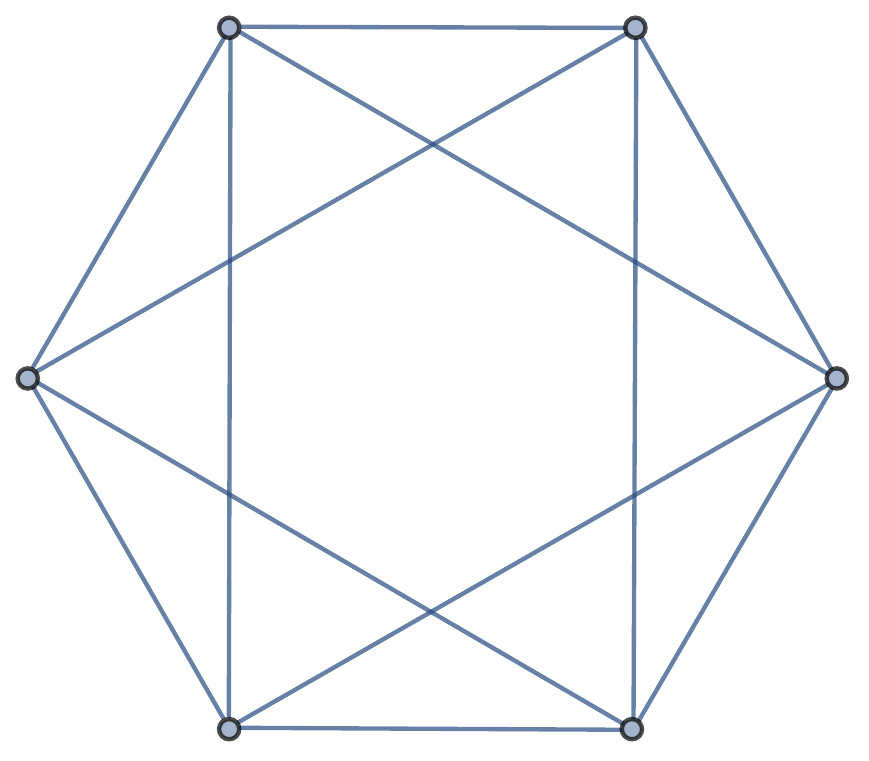}}}	 &	\multirow{2}{*}{Octahedron}	 &   & $4$ & {\tt{$3\times$Ising}},\, {\tt{O(3)}},\,  {\tt{Cubic}$_3$} \\
		& & & & 3 & {\tt{$3\times$Tri-Ising}},\, {\tt{Tri-O(3)}},\,  {\color{black}$\phi^6$-{\tt{Cubic}}$_3$} \\
& \\
		& \multirow{3}{*}{\parbox[c]{1em}{\includegraphics[scale=.1]{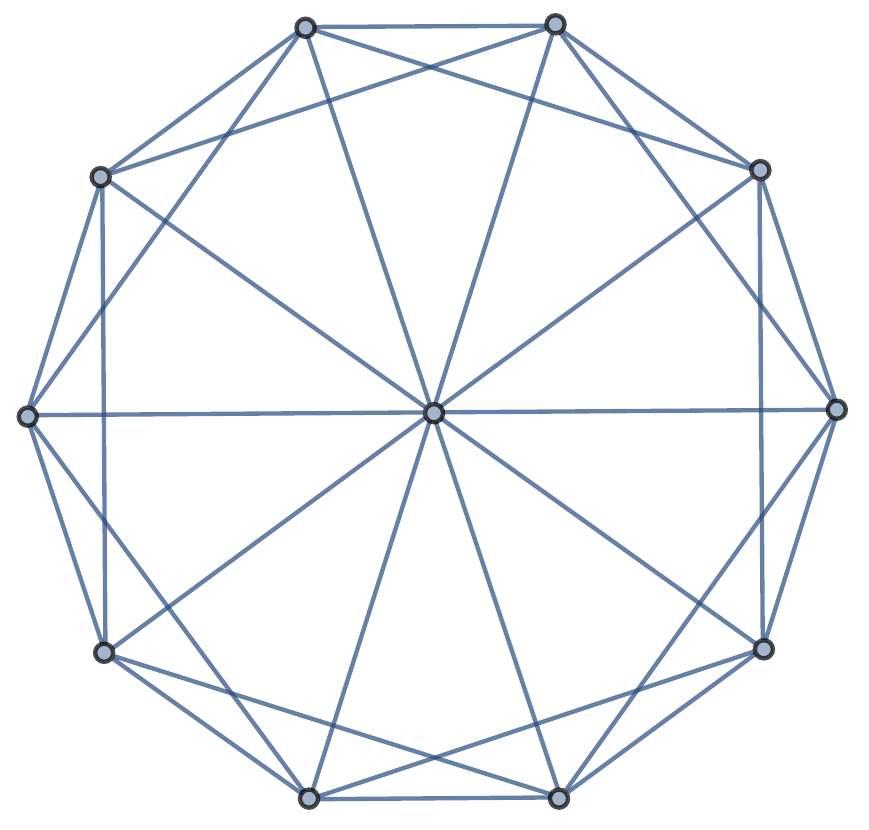}}} &	\multirow{3}{*}{Icosahedron} &	&	 3	& {\tt{Tri-O(3)}}  \\
		& &	&	 &	 $\nicefrac{8}{3}$	& {\color{black}{\tt{Tetra-O(3)}}}  \\
		& &	&	&	 $\nicefrac{5}{2}$	& {\color{black}{\tt{Penta-O(3)}}}, \, {\color{black}{\tt{Ico}$_{1\leq i \leq 2}$}}
		 \\
\\ \hline \\
\multirow{15}{*}{$N=4$}
		& \multirow{3}{*}{\parbox[c]{1em}{\includegraphics[scale=.1]{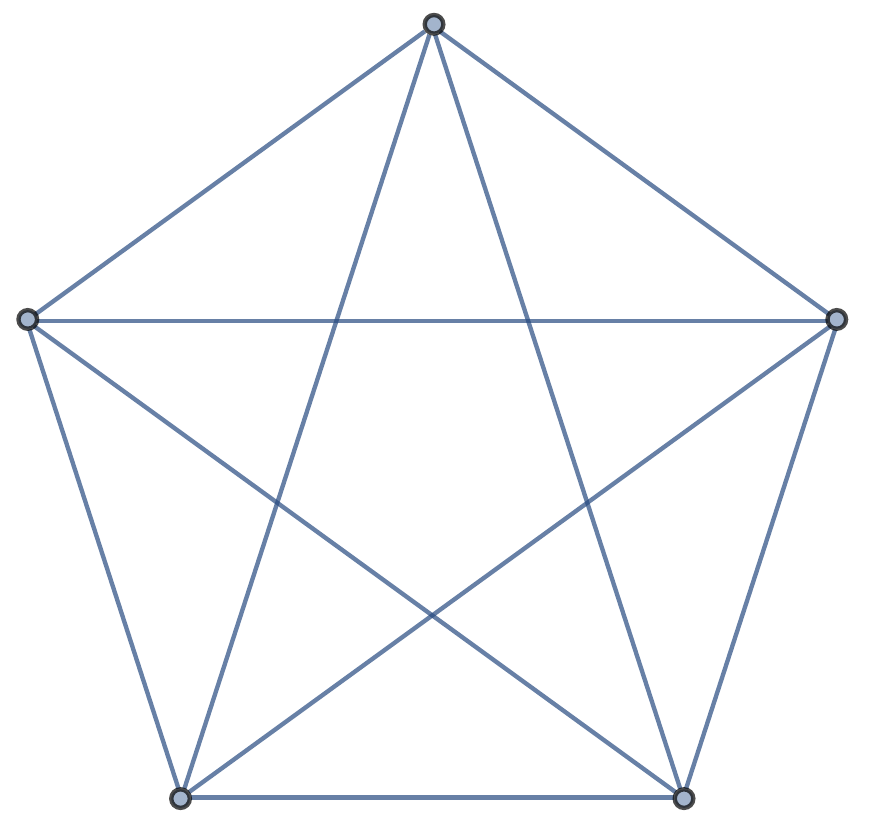}}}	 &	\multirow{3}{*}{5-cell}	 &	&	 $6$	& {\tt{No real FP}} \\
		&	&	&	 	&	 $4$	& {\tt{O(4)}}, \,  {\tt{Quartic-Potts}$_5$} \\
		&	&	&	&	 $\nicefrac{10}{3}$	& {\tt{No real FP}} \\
		& \\
		& \multirow{3}{*}{\parbox[c]{1em}{\includegraphics[scale=.1]{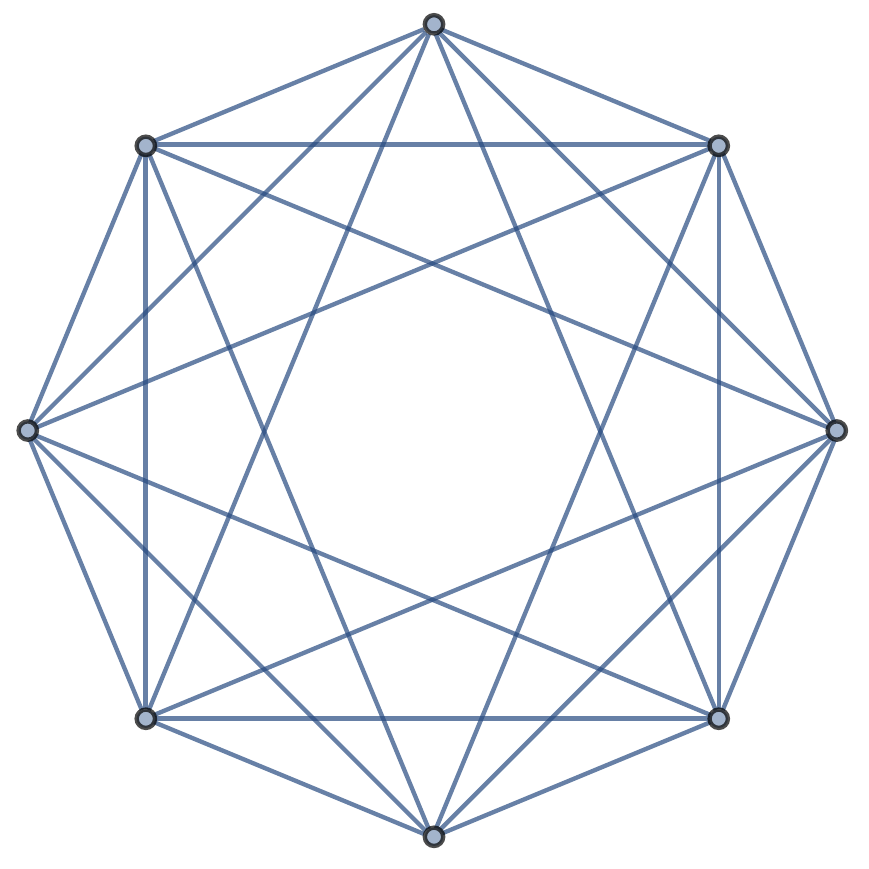}}}	 &	\multirow{3}{*}{$16$-cell}	 &	&	 $4$	& {\tt{$4\times$Ising}},\, {\tt{O(4)}} \\
		&	&	&	 	&	 $3$	& {\tt{$4\times$Tri-Ising}},\, {\tt{Tri-O(4)}},\,  {\color{black}$\phi^6$-{\tt{Cubic}}$_4$} \\
		&	&	&	&	 $\nicefrac{8}{3}$	&  {\tt{$4\times$Tetra-Ising}},\, {\color{black}{\tt{Tetra-O(4)}}},\,  {\color{black}$\phi^8$-{\tt{Cubic}}$_4$} \\
		& \\
		& \multirow{4}{*}{\parbox[c]{1em}{\includegraphics[scale=.1]{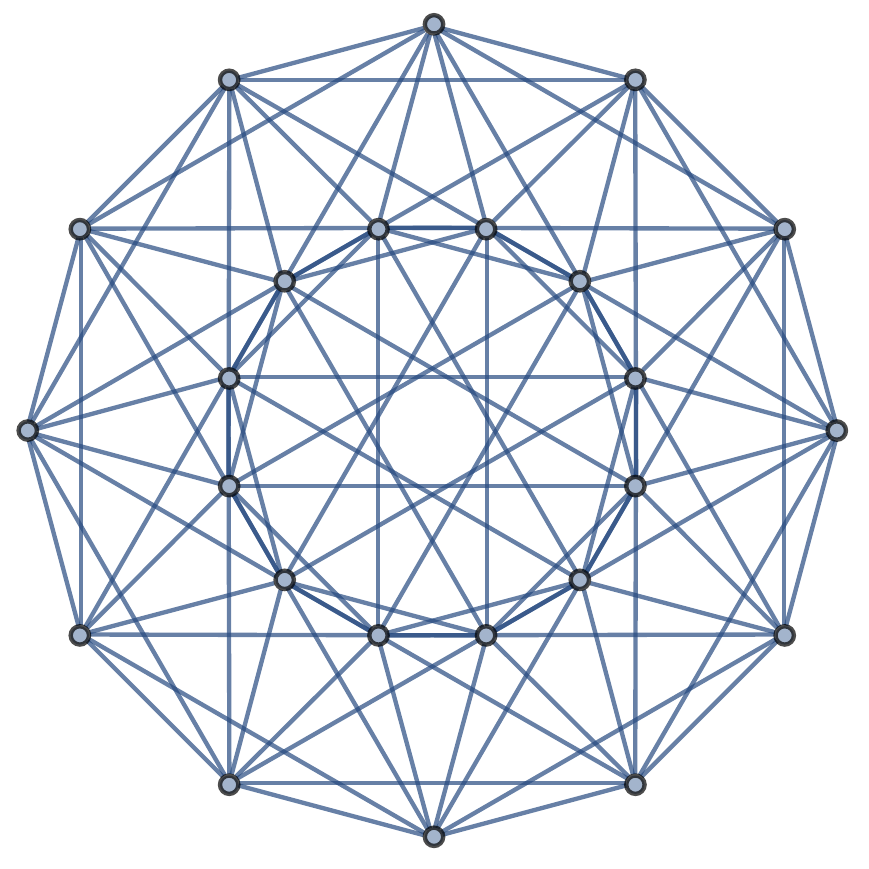}}} &	\multirow{4}{*}{$24$-cell} &	& $3$	& {\tt{Tri-O(4)}} \\
		& &	&		&	 $\nicefrac{8}{3}$	& {\color{black}{\tt{Tetra-O(4)}}}  \\
		& &	&	&	 $\nicefrac{5}{2}$	& {\color{black}{\tt{Penta-O(4)}}}, \, {\color{black}{\tt{24-cell}$_{1}$}} \\
		& &	&	&	 $\nicefrac{12}{5}$	& {\color{black}{\tt{Hexa-O(4)}}}, \, {\color{black}{\tt{24-cell}$_{1\leq i \leq 2}$}} \\
		& \\
		& \multirow{6}{*}{\parbox[c]{1em}{\includegraphics[scale=.1]{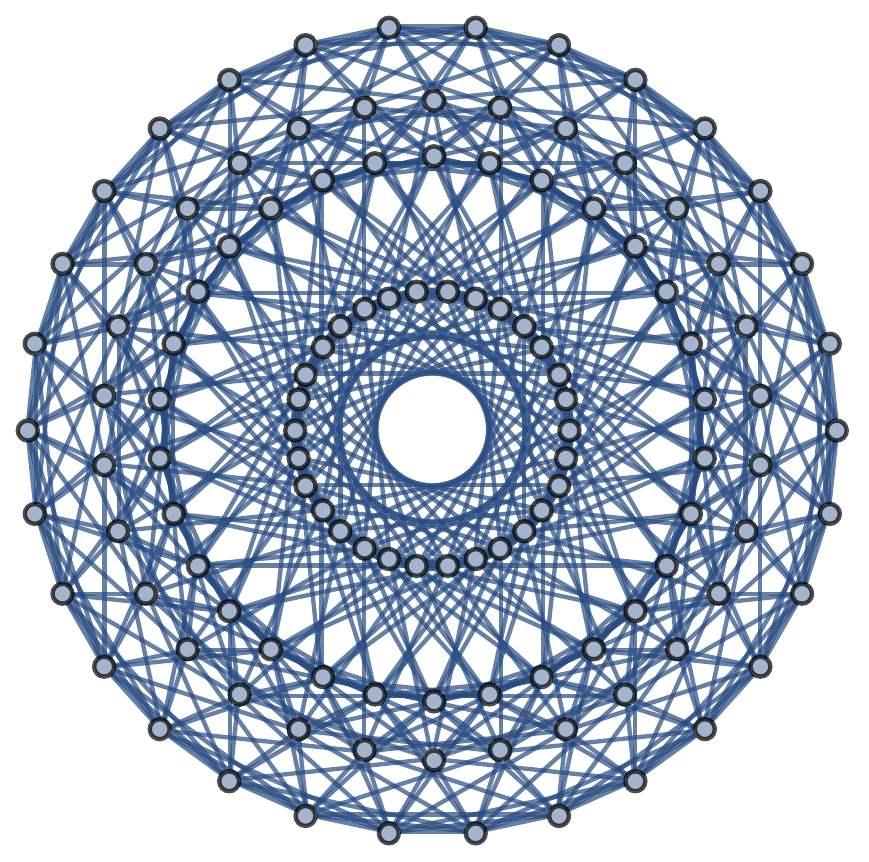}}} &	\multirow{6}{*}{$600$-cell} &	& \nicefrac{12}{5}	& {\color{black}\tt{Hexa-O(4)}} \\
		& & & &\vdots & \vdots\\
		& &	&	 	&	 \nicefrac{20}{9}& {\color{black}\tt{Deca-O(4)}} \\
		& & & &\vdots & \vdots\\
		& &	&	&	\nicefrac{ 15}{7}	&  {\color{black}\tt{Triaconta-O(4)}}  \\ \\
\end{tabular}
\caption{$N$-dimensional regular polytopes along with the upper critical dimensions $d_c$ around which the corresponding PFT can be studied in the $\epsilon$-expansion. For each polytope and $d_c$ we report the real FPs found in our analysis.\label{TFP}}
\end{ruledtabular}
\end{table}
%

\begin{table}[H]
\begin{center}
\begin{tabular}{cccccc|c|cc}
\hline\hline
	&&	{\tt Universality Class}	&&	$d_c$	&&	$\eta$	&&	$\nu$ \\\hline
&&&&&&&& \\
 \multirow{5}{*}{$N=1$} &&	{\tt Ising}	&&	4	&& $\frac{1}{54}\epsilon^2 $	&& $\frac{1}{2}+\frac{1}{12}\epsilon+\frac{7}{162}\epsilon^2 $ \\
&&&&&&&& \\
&&	{\tt Tri-Ising}	&&	3	&&	$\frac{1}{500}\epsilon^2$	&&	$\frac{1}{2}+\frac{1}{125}\epsilon^2$  \\
&&&&&&&&\\
&&	{\tt Tetra-Ising}	&&	$\frac{8}{3}$	&&	$\frac{9}{85750}\epsilon^2$	&&	$\frac{1}{2}+\frac{27}{68600}\epsilon^2$ \\
&&&&&&&& \\
\hline&&&&&&&&\\
&&	{\tt Potts}$_3$	&&	6	&& 	$\frac{1}{3}\epsilon $		&& $\frac{1}{2}-\frac{5}{12}\epsilon $ \\
&&&&&&&& \\
\multirow{7}{*}{$N=2$}&&	{\tt O(2)}	&&	4	&& $\frac{1}{50}\epsilon^2$	&&	$\frac{1}{2}+\frac{1}{10}\epsilon+\frac{11}{200}\epsilon^2$ \\
&&&&&&&& \\
&&	{\tt {\color{black}Pentagon}} 	&&	$\frac{10}{3}$	&&	$\frac{3}{5}\epsilon$	&&	$\frac{1}{2}+\frac{3}{20}\epsilon$ \\
&&&&&&&& \\
&&	{\tt Tri-O(2)}	&&	$3$	&&	$\frac{1}{392}\epsilon^2$	&&	$\frac{1}{2}+\frac{1}{98}\epsilon^2$ \\
&&&&&&&& \\
&& {\tt {\color{black}Heptagon}}	&&	$\frac{14}{5}$	&&	$\frac{10}{7}\epsilon$	&&	$\frac{1}{2}+\frac{5}{14}\epsilon$  \\
&&&&&&&& \\
&&	{\tt Tetra-O(2)}	&&	$\frac{8}{3}$	&&	$\frac{9}{59858}\epsilon ^2$	&&	$\frac{1}{2}+\frac{135}{239432} \epsilon ^2$ \\
&&&&&&&& \\
\hline&&&&&&&&\\
\multirow{9}{*}{$N=3$}	&&	{\tt O(3)}	&&	4	&&	$\frac{5}{242}\epsilon^2$	&&	$\frac{1}{2}+\frac{5}{44}\epsilon+\frac{345}{5324}\epsilon^2$ \\
&&&&&&&& \\
&&	{\tt Cubic}$_3$	&&	4	&&	$\frac{5 }{243}\epsilon^2$	&&	$\frac{1}{2}+\frac{1}{9}\epsilon+\frac{599}{8748}\epsilon^2$\\
&&&&&&&& \\
&&	{\tt Tri-O(3)}	&&	3	&&	$\frac{35}{11532}\epsilon^2$	&&	$\frac{1}{2}+\frac{35}{2883}\epsilon^2$\\
&&&&&&&& \\
&& {\color{black}$\phi^6$-{\tt{Cubic}}$_3$} 	&&	3	&&	$0.00261529\,\epsilon^2$	&&	$\frac{1}{2}+0.0104612\,\epsilon^2$\\
&&&&&&&& \\
&&	{\color{black}{\tt Tetra-O(3)}}	&&	$\frac{8}{3}$	&&	$\frac{945}{4798802}\epsilon^2$	&&	$\frac{1}{2}+\frac{14175}{19195208}\epsilon^2$ \\
&&&&&&&&\\
\hline&&&&&&&&\\
\multirow{11}{*}{$N=4$}	&&	{\tt O(4)} &&	$4$	&&	$\frac{1}{48}\epsilon^2$	&&	$\frac{1}{2}+\frac{1}{8}\epsilon+\frac{7}{96}\epsilon^2$ \\
&&&&&&&& \\
&& {\tt Quartic-Potts}$_5$\;	&&	$4$	&&	$\frac{55}{2646}\epsilon^2$	&&	$\frac{1}{2}+\frac{5}{42}\epsilon+\frac{22465}{222264}\epsilon^2$\\
&&&&&&&& \\
&&	{\tt Tri-O(4)}	&&	$3$	&& $\frac{1}{289}\epsilon^2$	&&	$\frac{1}{2}+\frac{4}{289}\epsilon^2$  \\
&&&&&&&& \\
&& {\color{black}$\phi^6$-{\tt{Cubic}}$_4$}	&&	$3$	&& $0.00322216\,\epsilon^2$ && $\frac{1}{2}+0.0128886\,\epsilon^2$  \\
&&&&&&&& \\
&& {\color{black}{\tt Tetra-O(4)}} &&$\frac{8}{3}$&& $\frac{9}{36980}\epsilon^2$ && $\frac{1}{2}+\frac{27}{29584}\epsilon^2$   \\
&&&&&&&& \\
&& {\color{black}$\phi^8$-{\tt{Cubic}}$_4$\;} &&$\frac{8}{3}$&& $\,0.000196765\,\epsilon^2$ && $\frac{1}{2}+0.000737867\,\epsilon^2$ \\
&&&&&&&&\\	
\hline
\hline
\end{tabular}
\end{center}
\caption{Critical exponents $\eta$ and $\nu$ for the universality classes with $d_c=6,4,\nicefrac{10}{3},3,\nicefrac{14}{5},\nicefrac{8}{3}$ (for which we know both of them) ordered by the number of field components.\label{TCE}} 
\end{table}

\section{Conclusion and Outlook}\label{sec:Conclusion}
In this paper we systematically analysed the critical behavior of Platonic Field Theories (PFTs) within the $\epsilon$-expansion.
We devised a method to construct invariant polynomials w.r.t the discrete symmetry groups of the regular polytopes, in terms of which we expressed the first $N$ independent invariants by increasing polynomial order.  
Since the upper critical dimensions the corresponding PFTs entail are generally non-integer (though still rational), we derived the relative  novel RG flow  by generalising the single component beta functionals $\beta_V$ and $\beta_Z$ to their multicomponent counterparts in all the relevant $d_c$ considered. New results in this respect regard $d_c = \nicefrac{14}{5},\nicefrac{8}{3},\nicefrac{5}{2},\nicefrac{12}{5}$ for which we reported the corresponding beta functionals in the main text and in Appendix \ref{Appendix:BetaFunctionals}.

A very interesting result of this analysis regards  a new candidate universality class in $d=3$ dimensions with the symmetry group $\mathbb{D}_5$ of the Pentagon. 
Validating its existence and measuring its critical properties by other methods surely deserves attention. Numerical Monte Carlo investigations are currently being
pursued in this direction \cite{rbaz2}.
Moreover being the upper critical dimension very close to three ($\epsilon=\nicefrac{1}{3}$) it  would be an ideal testing ground for the FPRG.
It would also be desirable to have an accurate estimate of the critical exponents  of this
universality class by means of CFT  bootstrap methods in terms of which hyper-Tetrahedral and hyper-Cubic theories have recently been analysed \cite{Stergiou2018}. 
Other interesting results  concern new Icosahedron fixed points in $d<3$ as well as the fixed points of the $24$-Cell.
As a by product of the present analysis we found many multi-critical $O(N)$ and $\phi^n$-Cubic universality classes.

Since the recent renewed interest in the multi-critical $O(N)$-models \cite{bertrand1, bertrand2}, future perspectives regard the analogous analysis of the multi-critical behavior of Cubic theories.
We also notice that the universality classes with $d_c<3$ may correspond to some novel unitary $2d$ CFTs with discrete global symmetry and of central charge $c>1$; these theories are likely to be  irrational CFTs and can be studied with numerical conformal bootstrap methods \cite{Poland:2018epd}. 
As it has been studied in \cite{CodelloScaling} for $\mathbb{Z}_2$ scalar theories, it  would also be  interesting to systematically analyse polygons, in particular with respect to the $d\to2$ limit where we expect a countable infinity of FPs corresponding to para-fermionic CFTs \cite{Fateev}.

Finally further studies can be directed to the application of the formalism to the study of those field theories characterised by the discrete symmetry group of a general geometrical object in a $N$ component space.

\section{Acknowledgments}
We would like to thank S. Caracciolo, G. Delfino and S. Rychkov for useful suggestions and comments.
\newpage
\appendix

\section{Analytical Details}\label{Appendix:AnalyticalDetails}

\subsection{\normalsize $N=2$}

\subsubsection*{ \normalsize {\color{blue} {\rm Triangle} $\{ 3 \}$ }}
The Triangle $\{3\}$ $\mathbb{D}_3$ symmetry is encoded in the following two invariants 
\begin{eqnarray}  
\rho_{\{3\}} &=& \frac{3}{2} \left(\phi _1^2+\phi _2^2\right) \,,\label{rhotriangle}  \\
\tau_{\{3\}} &=& \frac{3}{4} \phi _2 \left(\phi _2^2-3 \phi _1^2\right)\,.\label{tautriangle}
\end{eqnarray}
Since the non-trivial invariant polynomial $\tau_{\{3\}}$ is of order $k=3$, we study the Triangle in $d_c=6$ and therefore we consider the following marginal potential
\begin{equation}
U(\tau)=\frac{1}{3!}X\,  \tau_{\{3\}}\,.
\end{equation}
The beta function $\beta_X$ and the anomalous dimension $\eta$ are obtained from the general formulae \eqref{Betasd=6} and they read
\begin{eqnarray}  
\beta_X & = &-\frac{1}{2}\epsilon X + \frac{9}{32}X^3 \,, \\
\eta & = & \frac{3}{16} X^2\,.
\end{eqnarray}
We find that the universality class associated to the Triangle is the well known ${\tt{Potts}}_3$.

\subsubsection*{ \normalsize {\color{blue} {\rm Square} $\{ 4 \}$ }}

The two invariants for the Square $\{4\}$ $\mathbb{D}_4$ symmetry are
\begin{eqnarray}  
\rho_{\{4\}} &=& 2 \left(\phi _1^2+\phi _2^2\right)\,, \label{rhosquare}  \\
\tau_{\{4\}} &=& \phi _1^4+6 \phi _2^2 \phi _1^2+\phi _2^4 \,.\label{tausquare}
\end{eqnarray}
We immediately note that $\tau_{\{4\}}$ can be related to $\phi _1^4 + \phi _2^4$ by a field redefinition allowed by the ${\tt{O(2)}}$-symmetry (in fact one can check that {\tt Cubic}$_2$ = {\tt Ising} \cite{Osborn2018}) and, given that $\tau_{\{4\}}$ is of polynomial order $k=4$, we study the Square in $d_c=4$. The corresponding marginal potential reads
\begin{equation}
U(\rho,\tau) = \frac{1}{4!} ( X\, \rho_{\{4\}}^2+ Y \, \tau_{\{4\}} )\,.
\end{equation} 
In terms of the general formulae \eqref{BBetasd=4}, the beta functions and the anomalous dimension read
 \begin{eqnarray}  
\beta_X &=& - \epsilon X+\frac{40 }{3}X^2 +4 X Y\!-\frac{1024}{9}X^3 \!-64 X^2 Y\!-8X Y^2 \,, \label{betaXsquare}\\
\beta_Y &=&  - \epsilon Y\!+ 6 Y^2+16 X Y\!-\frac{512}{3}X^2 Y \!-128X Y^2-24 Y^3 \,, \label{betaYsquare}\\
\eta&=& \frac{32}{9} X^2+\frac{8}{3} X Y+\frac{2}{3} Y^2\,.
\end{eqnarray}
The LO fixed point potentials are
\begin{eqnarray*}
V(\phi_1,\phi_2) &=& \frac{\epsilon}{144}\left(\phi _1^4+6 \phi _2^2 \phi _1^2+\phi _2^4\right)\,,  \\
V(\phi_1,\phi_2) &=& \frac{\epsilon}{80} \left(\phi _1^2+\phi _2^2\right){}^2\,, \\
V(\phi_1,\phi_2) &=& \frac{\epsilon}{72} \left(\phi _1^4+\phi _2^4\right)   \,.
\end{eqnarray*}
The first and last potentials represent two copies of {\tt Ising} related by the aforementioned field redefinition, while the middle one is the {\tt O(2)} class.
The computation of the critical exponents at NLO confirms this picture.

\subsubsection*{ \normalsize {\color{blue} {\rm{Pentagon}} $\{ 5 \}$ }}

The interesting $\mathbb{D}_5$-symmetric Pentagon $\{5\}$ case can be studied  considering the following two invariant polynomials
\begin{eqnarray}  
\rho_{\{5\}} &=& \frac{5}{2} \left(\phi _1^2+\phi _2^2\right) \,, \label{rhopentagon}\\
\tau_{\{5\}} &=& \frac{5}{16} \left(\phi _2^5-10 \phi _1^2 \phi _2^3+5 \phi _1^4 \phi _2\right)\,.\label{taupentagon}
\end{eqnarray}
Since the invariant polynomial $\tau_{\{5\}}$ is of field order $k=5$, the corresponding upper critical dimension around which the $\epsilon$-expansion is performed is $d_c = \nicefrac{10}{3}$. Since $\rho$ and its powers are even in the fields, there is only one marginal coupling  and consequently the marginal potential reads
\begin{equation}
U(\tau) = \frac{1}{5!}  X\, \tau_{\{5\}}\,.
\end{equation} 
Beta functionals in $d_c=\nicefrac{10}{3}$  are given in Eq. \eqref{Betasd=10/3}.
Since the beta functional $\beta_V$ in $d_c = \nicefrac{10}{3}$ does not contain $V_{a_1a_2a_3a_4a_5}$, the beta function $\beta_X$ receives non-tree level contributions only from the anomalous dimension and it reads
\begin{equation}
\beta_X = -\frac{3}{2} \epsilon  X+ \frac{625}{384} X^3\,,
\end{equation}
with anomalous dimension given by
\begin{equation}
\eta = \frac{125}{192} X^2\,.
\end{equation}
The solution $X_* = \frac{24}{25}  \sqrt{\epsilon}$  defines the {\tt Pentagon} universality class with critical exponents reported in Table \ref{TCE}.

\subsubsection*{ \normalsize {\color{blue} {\rm{Hexagon}} $\{ 6 \}$ }}

The Hexagon $\{6\}$ dihedral symmetry $\mathbb{D}_6$ can be expressed in terms of the following two independent invariant polynomials 
\begin{eqnarray}  
\rho_{\{6\}} &=& 3 \left(\phi _1^2+\phi _2^2\right) \,, \label{rhohexagon}\\
\tau_{\{6\}} &=& \frac{3}{16} \left(11 \phi _1^6+15 \phi _2^2 \phi _1^4+45 \phi _2^4 \phi
   _1^2+9 \phi _2^6\right)\,.\label{tauhexagon}
\end{eqnarray}
Since $\tau_{\{6\}}$ is of polynomial order $k=6$, the corresponding upper critical dimension is $d_c=3$ and accordingly we consider the following marginal potential
\begin{equation}
U(\rho,\tau) = \frac{1}{6!}  (X\, \rho^3_{\{6\}} + Y\, \tau_{\{6\}})\,.
\end{equation} 
The LO beta functions and the anomalous dimension can be obtained from the general formulae \eqref{BBetasd=3}; the result is as follows
\begin{eqnarray}  
\beta_X &=& -2 \epsilon\, X  + \frac{1008 }{5}X^2 +18 X Y +\frac{5 }{16} Y^2  \,,\\
\beta_Y &=&  -2 \epsilon\, Y +144 X Y+10 Y^2 \,,\\
\eta &=& \frac{648}{25} X^2+\frac{18}{5} X Y+\frac{11}{80} Y^2\,,
\end{eqnarray}
and they are enough to show that the only real fixed point belongs to the {\tt O(2)} class.
\subsubsection*{ \normalsize {\color{blue} {\rm{Heptagon}} $\{ 7 \}$ }}

The heptagonal symmetry $\mathbb{D}_7$ can be expressed in terms of the following two independent invariant polynomials
\begin{eqnarray}  
\rho_{\{7\}} &=& \frac{7}{2} \left( \phi _1^2+\phi _2^2 \right) \,,\label{rhoheptagon}\\
\tau_{\{7\}} &=& \frac{7}{64} \left(\phi _2^7 - 21 \phi _1^2 \phi _2^5 + 35 \phi _1^4 \phi _2^3 - 7 \phi _1^6 \phi _2\right)\,.\label{tauheptagon}
\end{eqnarray}
In this case the invariant $\tau_{\{7\}}$ is of field order $k=7$, so that the corresponding upper critical dimension is $d_c = \nicefrac{14}{5}$; we notice that the Heptagon is the first polygon for which $d_c<3$. As for the {\tt{Pentagon}}, there is only one marginal coupling since $\rho_{\{7\}}$ and its powers are even in the fields and therefore the corresponding marginal potential reads
\begin{equation}
U(\tau) = \frac{1}{7!}  X\, \tau_{\{7\}}\,.
\end{equation} 
The beta function $\beta_X$ receives a non vanishing contribution only from the anomalous dimension since the beta functional $\beta_V$ in Eq. \eqref{Betasd=14/5} is identically zero in this case. We have
\begin{align}
\beta_{X} &= -\frac{5}{2}\epsilon X + \frac{245}{73728} X^3 \,,\\
\eta &= \frac{35}{18432}X^2\,.
\end{align}
The solution $X_* = \frac{192}{7}  \sqrt{\epsilon}$ represents the {\tt Heptagon} FP with critical exponents reported in Table \ref{TCE}.

\subsubsection*{ \normalsize {\color{blue} {\rm{Octagon}} $\{ 8 \}$ }}

In the Octagon case the two independent invariants are
\begin{align}  
\rho_{\{8\}} &= 4 \left(\phi _1^2+\phi _2^2\right) \,, \label{rhooctagon}\\
\tau_{\{8\}} &= \frac{1}{8} \left(17 \phi _1^8+84 \phi _2^2 \phi _1^6+70 \phi _2^4 \phi
   _1^4+84 \phi _2^6 \phi _1^2+17 \phi _2^8\right)\,.\label{tauoctagon}
\end{align}
The invariant $\tau_{\{8\}}$ is of field order $k=8$ and we therefore analyse the theory in $d_c = \nicefrac{8}{3}$. As for the Hexagon and the Square, there are two marginal couplings and the corresponding marginal potential reads
\begin{equation}
U(\rho,\tau) = \frac{1}{8!} \left( X\, \rho^4_{\{8\}}+  Y\, \tau_{\{8\}}\right) \,.
\end{equation} 
The beta functions and anomalous dimension can both be extracted from Eq. \eqref{BBetasd=8/3} and they read
\begin{align}
\beta_{X} &= - 3\epsilon X+\frac{88576}{35} X^2+\frac{69}{2} X Y +\frac{455}{4096}  Y^2  \,,\\
\beta_{Y} &= -3\epsilon Y+ 1024 X Y+\frac{35}{4} Y^2 \,,\\
\eta &= \frac{131072}{1225} X^2+\frac{64}{35} X Y+\frac{9}{1120} Y^2\,.
\end{align}
The solution $\{ X_* = \frac{105 }{88576}\epsilon , Y_*=0 \}$ represents the {\tt{Tetra-O(2)}} universality class with exponents reported in Table \ref{TCE}.

\subsection{\normalsize $N=3$}
\subsubsection*{ \normalsize {\color{blue} {\rm{Tetrahedron}} $\{3,3\}$}}
The Tetrahedron (Potts$_4$)  $S_4$ symmetry has been widely analysed and can be studied in terms of the following polynomial invariants
\begin{eqnarray}  
\rho_{\{3,3\}} &=& \frac{4}{3} \left(\phi _1^2+\phi _2^2+\phi _3^2\right)\,, \label{rhotetra}\\
\tau_{\{3,3\}} &=& \frac{4}{9} \left(\sqrt{2} \phi _1^3-3 \phi _3 \phi _1^2-3 \sqrt{2} \phi _2^2
   \phi _1+2 \phi _3^3-3 \phi _2^2 \phi _3\right)\,, \label{tautetra}\\
\sigma_{\{3,3\}} & =& \frac{4}{27}\! \left(6 \phi _1^4-4 \sqrt{2} \phi _3 \phi _1^3+6\!\left(2 \phi_2^2+\phi _3^2\right)\!\phi _1^2+12 \sqrt{2} \phi _2^2 \phi _3 \phi _1+6
   \phi _2^4+7 \phi _3^4+6 \phi _2^2 \phi _3^2\right).\label{sigmatetra}
\end{eqnarray}
Since $\tau_{\{3,3\}}$ and $\sigma_{\{3,3\}}$ appear respectively at order $k=3$ and $k=4$, the upper critical dimensions in the Tetrahedron case are $d_c=6,4$. In $d_c=6$ we have only one marginal coupling
\begin{equation}
U(\tau) = \frac{1}{3!} X \,\tau_{\{3,3\}} \,,
\end{equation} 
and at LO the beta function $\beta_X$ and anomalous dimension can be obtained from Eq. \eqref{Betasd=6} as
\begin{eqnarray}
\beta_X &=& -\frac{1}{2}\epsilon X -\frac{8}{27} X^3 \,,\\
\eta &=& \frac{16}{81}  X^2\,,
\end{eqnarray} 
and it does not have any non-trivial real FP. In $d_c=4$ instead we have two marginal couplings and the marginal potential is given by
\begin{equation}
U(\rho,\tau) = \frac{1}{4!} (X \,\rho^2_{\{3,3\}} + Y \,\sigma_{\{3,3\}} )\,.
\end{equation} 
At NLO we find the following beta functions and anomalous dimension
\begin{eqnarray}
\beta_X &=&-\epsilon\, X -\frac{5888}{243} X^3-\frac{1408}{81} X^2 Y+\frac{176}{27} X^2-\frac{1136}{243} XY^2+\frac{8}{3} X Y - \frac{16}{27} Y^3+\frac{1}{3}Y^2 \,,\\
\beta_Y &=& -\epsilon \,Y - \frac{24832}{729} X^2 Y - \frac{2176}{81} X Y^2+\frac{64}{9} X Y - \frac{1328}{243} Y^3 + \frac{8}{3}Y^2 \,, \\
\eta &=& \frac{640}{729} X^2+\frac{64}{81} X Y+\frac{56}{243} Y^2\,.
\end{eqnarray} 
As already pointed out in Section \ref{UC}, by symmetry enhancement the universal content of the $d_c=4$ Tetrahedron is the same as the $d_c=4$ Cube one (see Table \ref{TFP} with corresponding critical exponents reported in Table \ref{TCE}).

\subsubsection*{ \normalsize {\color{blue} {\rm{Octahedron}} $\{3,4\}$ -  {\rm{Cube}} $\{4,3\}$ }}

The Octahedron-Cube $S_4 \times \mathbb{Z}_2$ symmetry can be easily expressed in the Octahedron basis, in terms of which the invariant polynomials assume a very simple form
\begin{align}  
\rho_{\{3,4\}} &= 2\, (\phi _1^2+\phi _2^2+\phi _3^2  ) \,,\label{rhoocta}\\
\tau_{\{3,4\}} &= 2\, (\phi _1^4+\phi _2^4+\phi _3^4  ) \,,\label{tauocta}\\
\sigma_{\{3,4\}} & = 2\, (\phi _1^6+\phi _2^6+\phi _3^6  )\,.\label{sigmaocta}
\end{align}
The duality map that relates the Octahedron invariants to the Cube ones is given in Eq. \eqref{eq:octatocube}. The non-trivial invariant polynomials $\tau_{\{3,4\}}$ and $\sigma_{\{3,4\}}$ are respectively of order $k=4$ and $k=6$ and consequently the upper critical dimensions we consider in this case are $d_c = 4,3$.

In $d_c=4$ there are only two marginal couplings since $\sigma_{\{3,4\}}$ is irrelevant and the marginal potential reads
\begin{equation}
U(\rho,\tau) = \frac{1}{4!} (X \,\rho^2_{\{3,4\}} + Y \,\tau_{\{3,4\}} )\,,
\end{equation} 
with NLO beta functions and anomalous dimension computed from the general expressions \eqref{BBetasd=4}
\begin{eqnarray}  
\beta_X &=&-\epsilon\, X +\frac{44 }{3}X^2+4 X Y-\frac{20 }{3}X Y^2-\frac{368 }{3}X^3-\frac{176}{3} X^2 Y  \,,\\
\beta_Y &=& -\epsilon\,Y+6 Y^2+16 X Y-\frac{1552}{9} X^2 Y-\frac{368}{3} X Y^2-\frac{68}{3} Y^3 \,,\\
\eta &=& \frac{40}{9} X^2 + \frac{8}{3}  X Y +\frac{2}{3} Y^2\,.
\end{eqnarray}
This systems has three non-trivial fixed points that correspond to three copies of {\tt Ising}, {\tt O(3)} and {\tt Cubic}$_3$ universality classes.
Their GL potentials are, respectively,
\begin{eqnarray*}
V_{\tt 3\times Ising} &=& \frac{ \epsilon }{72} (\phi _1^4+\phi _2^4+\phi _3^4 )  \,,\\
V_{\tt O(3)} &=& \frac{\epsilon}{88} \left(\phi _1^2+\phi _2^2 +\phi _3^2\right)^2   \,,\\
V_{{\tt Cubic}_3} &=& \frac{\epsilon}{72} \left(\phi _1^2+\phi _2^2+\phi _3^2\right){}^2 -\frac{\epsilon}{216}
   \left(\phi _1^4+\phi _2^4+\phi _3^4\right) \,.
\end{eqnarray*}
The critical exponents are reported in Table \ref{TCE}.

In $d_c=3$ instead there are three marginal couplings 
\begin{equation}
U(\rho,\tau,\sigma) = \frac{1}{6!} ( X\, \rho^3_{\{3,4\}} + Y \, \rho_{\{3,4\}} \tau_{\{3,4\}} + Z \, \sigma_{\{3,4\}} )\,,
\end{equation} 
and we give here the LO beta functions and anomalous dimension
 \begin{eqnarray}  
\beta_X &=& -2 \epsilon\, X  +\frac{992}{15} X^2+\frac{64}{5} X Y +\frac{8}{15} Y^2 \,, \\
\beta_Y &=& -2 \epsilon\, Y  +\frac{208}{15} Y^2+\frac{448}{5} X Y+32 X Z+\frac{16}{3} Y Z \,,\\
\beta_Z &=& -2 \epsilon\, Z  +\frac{40}{3} Z^2+\frac{128}{3} X Z+\frac{224}{9} Y^2+\frac{128}{3} Y Z \,,\\
\eta &=& \frac{448}{135} X^2+\frac{448 X Y}{225} X Y+\frac{32 X Z}{45} X Z+\frac{272}{675} Y^2+\frac{16}{45} YZ + \frac{4}{45} Z^2\,.
\end{eqnarray}
The non-trivial FPs turn out to be, as expected, the tri-critical version of the three FPs in $d_c=4$. Their critical exponents are reported in Table \ref{TCE}.

\subsubsection*{ \normalsize {\color{blue}  {\rm{Icosahedron}} $\{3,5\}$ -  {\rm{Dodecahedron}} $\{5,3\}$ }}

In the Icosahedron basis, the $A_5\times\mathbb{Z}_2$ symmetric independent invariant polynomials read
\begin{eqnarray}  
\rho_{\{3,5\}} &=& 4 (\phi _1^2+\phi _2^2+\phi _3^2 ) \,,\label{rhoico} \\
\tau_{\{3,5\}} &=& \frac{4}{25} (10 \phi _1^6+6 \phi _3 \phi _1^5+15 \left(2 \phi _2^2+3
   \phi _3^2\right) \phi _1^4-60 \phi _2^2 \phi _3 \phi _1^3\nonumber\\&&+15 \left(2
   \phi _2^4+6 \phi _3^2 \phi _2^2+\phi _3^4\right) \phi _1^2+30 \phi _2^4
   \phi _3 \phi _1+10 \phi _2^6+13 \phi _3^6+15 \phi _2^2 \phi _3^4+45
   \phi _2^4 \phi _3^2) \,, \label{tauico} \\
\sigma_{\{3,5\}} & =& \frac{4}{625} (127 \phi _1^{10}+360 \phi _3 \phi _1^9+45 \left(13 \phi
   _2^2+35 \phi _3^2\right) \phi _1^8-120 \left(24 \phi _2^2 \phi _3-7 \phi
   _3^3\right) \phi _1^7\nonumber\\&&+210 \left(7 \phi _2^4+30 \phi _3^2 \phi _2^2+10 \phi
   _3^4\right) \phi _1^6-252 \left(-\phi _3^5+30 \phi _2^2 \phi _3^3+20 \phi
   _2^4 \phi _3\right) \phi _1^5\nonumber\\&&+210 \left(5 \phi _2^6+45 \phi _3^2 \phi
   _2^4+30 \phi _3^4 \phi _2^2+3 \phi _3^6\right) \phi _1^4-840 \left(3 \phi
   _2^2 \phi _3^5+5 \phi _2^4 \phi _3^3\right) \phi _1^3\nonumber\\&&+45 \left(15 \phi
   _2^8+140 \phi _3^2 \phi _2^6+140 \phi _3^4 \phi _2^4+28 \phi _3^6 \phi
   _2^2+\phi _3^8\right) \phi _1^2+60 \left(30 \phi _3 \phi _2^8+70 \phi _3^3
   \phi _2^6+21 \phi _3^5 \phi _2^4\right) \phi _1\nonumber\\&&+125 \phi _2^{10}+313
   \phi _3^{10}+45 \phi _2^2 \phi _3^8+630 \phi _2^4 \phi _3^6+2100 \phi
   _2^6 \phi _3^4+1575 \phi _2^8 \phi _3^2) \,, \label{sigmaico}
\end{eqnarray}
which can be expressed, by duality, in the Dodecahedron basis in terms of the following map
\begin{align}  
\rho_{\{5,3\}} =& \frac{5}{3}\,\rho_{\{3,5\}}\,, \nonumber \\
\tau_{\{5,3\}} =& \frac{5}{72}\,\rho^3_{\{3,5\}} - \frac{25}{27}\,\tau_{\{3,5\}} \,, \nonumber \\
\sigma_{\{5,3\}} =& \frac{35}{2592}\,\rho^5_{\{3,5\}} - \frac{175}{324}\,\rho^2_{\{3,5\}}\tau_{\{3,5\}} + \frac{625}{243}\,\sigma_{\{3,5\}}\,.
\end{align}
Since the field power of $\tau_{\{3,5\}}$ and $\sigma_{\{3,5\}}$ are respectively $k=6$ and $k=10$ the interesting upper critical dimensions are $d_c = 3, \nicefrac{8}{3}, \nicefrac{5}{2}$. 

In $d_c=3$ the marginal potential is
\begin{equation}
U(\rho,\tau) = \frac{1}{6!} \left(X\, \rho^3_{\{3,5\}} + Y\, \tau_{\{3,5\}} \right)\,,
\end{equation} 
and the corresponding LO beta functions, computed from Eq. \eqref{BBetasd=3}, read
\begin{eqnarray}  
\beta_X &=&-2\epsilon\, X  +\frac{7936 }{15}X^2+\frac{96 }{5}X Y +\frac{3}{25} Y^2 \,,\\
\beta_Y &=&  -2\epsilon\, Y +\frac{32 }{3}Y^2 +\frac{1024}{3} X Y \,,
\end{eqnarray}
with anomalous dimension
\begin{equation}
\eta = \frac{28672 }{135}X^2 +\frac{512 }{45}X Y+\frac{208 }{1125}Y^2\,.
\end{equation} 
This system exhibits no other real fixed point than {\tt Tri-O(3)} for which the critical exponents are given in Table \ref{TCE}.

Also in $d_c=\nicefrac{8}{3}$ we find only {\tt Tetra-O(3)}, whose critical exponents are reported in Table \ref{TCE}.
To find a real icosahedral fixed point we have to shift to the third possible upper critical dimension which is $d_c= \nicefrac{5}{2}$ for which the marginal potential assumes the following form
\begin{equation}\label{icopot5/2}
U(\rho,\tau,\sigma) = \frac{1}{10!} \left(X\, \rho^5_{\{3,5\}} + Y\, \rho^2_{\{3,5\}} \tau_{\{3,5\}} + Z\, \sigma_{\{3,5\}} \right)\,.
\end{equation} 
The LO beta function system in this case reads
\begin{eqnarray}  
\beta_X &=&-4\epsilon X+\frac{10381312}{945} X^2+\frac{3968}{15} X Y - \frac{896}{45} X Z +\frac{19601}{23625} Y^2 - \frac{1817}{4500} Y Z + \frac{203}{200000} Z^2 \,, \nonumber \\
\beta_Y &=& -4\epsilon Y + \frac{11429888}{945} X Y + \frac{68864}{45} X Z + \frac{232928}{945} Y^2 + \frac{7024}{225} Y Z - \frac{7}{250} Z^2 \,,\nonumber \\
\beta_Z &=& -4\epsilon Z + \frac{32768}{15} X Z + \frac{13568}{21} Y^2 + \frac{3776}{15} Y Z + \frac{84}{5} Z^2\,,
\end{eqnarray}
with anomalous dimension
\begin{equation}  
\eta =\frac{2883584 }{42525}X^2+\frac{360448 }{99225}X Y+\frac{2048 }{14175}X Z+\frac{381952  }{7441875} Y^2
+\frac{1664 }{354375}Y Z+\frac{1252}{8859375} Z^2\,.
\end{equation} 
Apart from the {\tt Penta-O(3)} FP (see Table \ref{TCE}) there are two pure icosahedral real FPs.

\subsection{\normalsize $N=4$}
\subsubsection*{ \normalsize {\color{blue} {\rm{5-cell}} $\{3,3,3\}$}}
The $S_5$ symmetric $5$-cell can be studied in terms of the following polynomial invariants
{\small
\begin{align}  
\rho_{\{3,3,3\}} &= \frac{5}{4} \left(\phi _1^2+\phi _2^2+\phi _3^2+\phi _4^2\right) \,,\label{rho5cell}\\
\tau_{\{3,3,3\}} &= \frac{15 \left(\phi _3 \phi _1^2+\left(\phi _3^2+\left(2 \phi _2-\phi
   _4\right) \phi _4\right) \phi _1-\phi _2 \phi _3 \left(\phi _2+2 \phi
   _4\right)\right)}{8 \sqrt{2}} \,,\label{tau5cell}\\
\sigma_{\{3,3,3\}} &= \frac{5}{32} \left(3 \phi _1^4+4 \phi _3 \phi _1^3+6 \left(\phi _2^2-2 \phi
   _4 \phi _2+2 \left(\phi _3^2+\phi _4^2\right)\right) \phi _1^2+4 \phi _3
   \left(-3 \phi _2^2+\phi _3^2-3 \phi _4^2\right) \phi _1 \right. \nonumber\\
   & + 3\left. \phi _2^4+3
   \left(\phi _3^2+\phi _4^2\right){}^2+4 \phi _2^3 \phi _4+12 \phi _2^2
   \left(\phi _3^2+\phi _4^2\right)+4 \phi _2 \left(3 \phi _3^2 \phi
   _4-\phi _4^3\right)\right) \,,\label{sigma5cell}\\
   \omega_{\{3,3,3\}} &= \frac{5}{64 \sqrt{2}} \left(\phi _1^5+20 \phi _3 \phi _1^4-10 \left(\phi _2^2-4 \phi _4
   \phi _2-3 \phi _3^2+3 \phi _4^2\right) \phi _1^3+30 \left(\phi
   _3^3+\phi _4 \left(\phi _4-2 \phi _2\right) \phi _3\right) \phi _1^2\right.\nonumber \\
   & +\left.5
   \left(\phi _2^4+8 \phi _4 \phi _2^3+6 \left(\phi _3^2-\phi _4^2\right)
   \phi _2^2+12 \phi _4 \left(\phi _3^2+\phi _4^2\right) \phi _2+4
   \left(\phi _3^4-\phi _4^4\right)\right) \phi _1\right. \nonumber\\
   &+ \left. \phi _3 \left(-20 \phi
   _2^4-60 \phi _4 \phi _2^3-30 \left(\phi _3^2+\phi _4^2\right) \phi _2^2-40
   \phi _4 \left(\phi _3^2+\phi _4^2\right) \phi _2+\phi _3^4+5 \phi
   _4^4-10 \phi _3^2 \phi _4^2\right)\right)\,.\label{omega5cell}
\end{align}
}
The Invariants $\tau_{\{3,3,3\}}$, $\sigma_{\{3,3,3\}}$ and $\omega_{\{3,3,3\}}$ appear respectively at order $k=3,4,5$ and therefore we study the critical behaviour of the $5$-cell at the upper critical dimensions $d_c=6,4,\nicefrac{10}{3}$.

In $d_c=6$ we have only one marginal coupling and the corresponding potential reads
\begin{equation}
U(\tau) = \frac{1}{6} X\, \tau_{\{3,3,3\}}\,,
\end{equation} 
with the corresponding LO beta function $\beta_X$ and anomalous dimension $\eta$ given by
\begin{eqnarray}
\beta_X &=& - \frac{1}{2}\epsilon \,X -\frac{125}{256}  X^3 \,,\\
\eta &=& \frac{25}{128} X^2\,.
\end{eqnarray} 
In $d_c=4$ instead we have two marginal couplings
\begin{equation}
U(\rho,\tau)=\frac{1}{4!} \left(  X\, \rho^2_{\{3,3,3\}} + Y\,\sigma_{\{3,3,3\}}\right)\,,
\end{equation} 
and at NLO we find the following system of beta functions along with the corresponding anomalous dimension
\begin{align}
\beta_X &= - \epsilon X - \frac{8125}{384} X^3 - \frac{1375}{96} X^2 Y + \frac{25}{4} X^2 - \frac{2525}{768} X Y^2 + \frac{5}{2} X Y - \frac{45}{128} Y^3 + \frac{3}{16} Y^2 \,, \\
\beta_Y &= -\epsilon Y - \frac{10625}{384} X^2 Y - \frac{4625}{192} X Y^2 + \frac{25}{4} X Y - \frac{4175}{768} Y^3 + \frac{45}{16} Y^2 \,,\\
\eta &=\frac{625}{768} X^2 + \frac{125}{192} X Y + \frac{325}{1536} Y^2\,.
\end{align} 

Finally in $d_c=\nicefrac{10}{3}$ we also have two marginal couplings and the potential reads
\begin{equation}
U(\rho, \tau, \omega) = \frac{1}{5!} \left( X \,\rho_{\{3,3,3\}}  \tau_{\{3,3,3\}} + Y \,\omega_{\{3,3,3\}}  \right)\,,
\end{equation} 
and we find the following LO beta functions $\beta_X$, $\beta_Y$ and anomalous dimension $\eta$
\begin{align}
\beta_X &= -\frac{3}{2} \epsilon X - \frac{6525}{4096} X^3 + \frac{17325}{2048} X^2 Y + \frac{98775 }{8192}X Y^2 + \frac{8025}{2048} Y^3 \,,\\
\beta_Y &= -\frac{3}{2}\epsilon Y - \frac{138375}{1024} X^3 - \frac{2948625}{8192} X^2 Y - \frac{640125}{2048} X Y^2 - \frac{732675}{8192} Y^3 \,,\\
\eta &= \frac{1125}{4096} X^2+\frac{225}{512} X Y+\frac{765}{4096} Y^2\,.
\end{align} 

\subsubsection*{ \normalsize {\color{blue} {\rm{16-Cell}} $\{3, 3, 4\}$   - {\rm{8-Cell}} $\{4, 3, 3\}$ }}

The four polynomial invariants in the $16$-cell basis are very simple and read
\begin{align}  
\rho_{\{3, 3, 4\}} &= 2 \left(\phi _1^2+\phi _2^2+\phi _3^2+\phi _4^2\right) \,,\label{rho16} \\
\tau_{\{3, 3, 4\}} &= 2 \left(\phi _1^4+\phi _2^4+\phi _3^4+\phi _4^4\right) \,,\label{tau16}\\
\sigma_{\{3, 3, 4\}} &= 2 \left(\phi _1^6+\phi _2^6+\phi _3^6+\phi _4^6\right) \,,\label{sigma16}\\
\omega_{\{3, 3, 4\}} &= 2 \left(\phi _1^8+\phi _2^8+\phi _3^8+\phi _4^8\right)\,. \label{omega16} 
\end{align}
The duality between the $16$-cell and the $8$-cell can be expressed in terms of the following map between the polynomial invariants in the two bases 
\begin{align}  
\rho_{\{4,3,3\}} &= 8\,\rho_{\{3, 3, 4\}}  \,, \nonumber \\
\tau_{\{4,3, 3\}} &= 12\,\rho^2_{\{3, 3, 4\}} - 16\, \tau\,_{\{3, 3, 4\}} \,,\nonumber \\
\sigma_{\{4,3, 3\}} &= 30\,\rho^3_{\{3, 3, 4\}} - 120\,\tau_{\{3, 3, 4\}}\rho_{\{3, 3, 4\}} + 128\,\sigma_{\{3, 3,4\}} \,, \nonumber \\
\omega_{\{4,3, 3\}} &= 105\,\rho^4_{\{3, 3, 4\}} - 840\,\rho^2_{\{3, 3, 4\}}\tau_{\{3, 3, 4\}}+ 1792\,\rho_{\{3, 3, 4\}}\sigma_{\{3, 3, 4\}} -2176\,\omega_{\{3, 3, 4\}}\,.
\end{align}
The invariants $\tau_{\{3,3,4\}}$, $\sigma_{\{3,3,4\}}$ and $\omega_{\{3,3,4\}}$ appear respectively at order $k=4,6,8$ so that the proper upper critical dimensions to study this theory are $d_c =4,3,\nicefrac{8}{3}$.

In $d_c=4$ the potential has two marginal couplings
\begin{equation}
U(\rho,\tau) = \frac{1}{4!} (X\, \rho^2_{\{3, 3, 4\}}  + Y\, \tau_{\{3, 3, 4\}}  )\,,
\end{equation} 
and we find the following NLO beta functions
\begin{align}  
\beta_X &=- \epsilon X +16 X^2+4 X Y-\frac{416}{3} X^3-\frac{176}{3} X^2 Y-\frac{20}{3} X Y^2 \,, \\
\beta_Y &= -\epsilon Y +6 Y^2-\frac{544}{3} X^2 Y-\frac{368}{3} X Y^2+16 X Y-\frac{68}{3} Y^3 \,,
\end{align}
with anomalous dimension
\begin{equation}  
\eta = \frac{16 }{3}X^2+\frac{8 }{3}X Y+\frac{2 }{3}Y^2\,.
\end{equation}

In $d_c=3$ we have three marginal couplings and the marginal potential reads
\begin{equation}
U(\rho,\tau,\sigma) = \frac{1}{6!} \left( X\, \rho^3_{\{3, 3, 4\}} + Y\,\rho^2_{\{3, 3, 4\}} \tau_{\{3, 3, 4\}}+ Z \, \sigma_{\{3, 3, 4\}} \right)\,,
\end{equation} 
and for simplicity we give only the LO beta funtions
\begin{eqnarray}  
\beta_X &=&-2 X \epsilon +\frac{1088}{15} X^2+\frac{64}{5} X Y +\frac{8}{15} Y^2 \,, \\
\beta_Y &=& -2 Y \epsilon + \frac{1408}{15} X Y+32 X Z  +\frac{208}{15} Y^2+\frac{16}{3} Y Z \,,  \\
\beta_Z &=&-2 Z \epsilon+ \frac{128}{3} X Z+\frac{128}{5} Y^2 +\frac{128}{3} Y Z+\frac{40}{3} Z^2 \,,
\end{eqnarray}
with
\begin{equation} 
\eta =
\frac{1024}{225} X^2+\frac{512}{225} X Y+\frac{32}{45} X Z +\frac{32}{75} Y^2+\frac{16}{45} YZ+\frac{4}{45} Z^2 \,.
\end{equation}

Finally in $d_c=\nicefrac{8}{3}$ the marginal potential has five couplings
\begin{equation}
U(\rho,\tau,\sigma,\omega) = \frac{1}{8!} \left(X\,\rho^4_{\{3, 3, 4\}} +Y\, \rho ^2_{\{3, 3, 4\}} \tau_{\{3, 3, 4\}} + Z\, \rho_{\{3, 3, 4\}}  \sigma_{\{3, 3, 4\}}  +W\, \tau ^2_{\{3, 3, 4\}}  +  T\,\omega_{\{3, 3, 4\}} \right)\,,
\end{equation} 
with LO beta functions given by
\begin{align}
\beta_X&= -3 \epsilon X+\frac{1376}{7} X^2+\frac{36}{35} W X+\frac{1032}{35} Y X+\frac{9}{7} Z X +\frac{37}{35} Y^2+\frac{3}{28} Y Z\,, \\
\beta_Y&= -3\epsilon  Y+\frac{148}{5} Y^2+\frac{1}{2}T Y+\frac{208}{35} W Y+\frac{1376}{5} X Y+\frac{68}{7} Z Y+\frac{45}{56} Z^2\nonumber\\&+6 T X+\frac{1824}{35} W X+\frac{3}{14} W Z+\frac{444}{7} X Z \,, \\
\beta_Z&= -3 Z
   \epsilon+\frac{64}{35} W^2+\frac{768}{7} X W+\frac{1952}{35} Y W+\frac{120}{7} Z W+\frac{7424}{105}Y^2+\frac{205}{14} Z^2 \nonumber \\
   &+96 T X+28 T Y+\frac{15}{2} T Z+\frac{1216}{7} X Z+80 Y Z  \,, \\
\beta_W&= -3 \epsilon  W+\frac{36}{7} W^2+T W+64 X W+\frac{944}{35} Y W+\frac{43}{7} Z W+\nonumber\\
& +\frac{2008}{105}Y^2+\frac{15}{56} Z^2+T Y+\frac{54}{7} Y Z \,, \\
\beta_T&= -3 \epsilon  T+\frac{35}{2} T^2+68 W T+64 X T+80 Y T+55 Z T+\frac{1968}{35} W^2+\frac{270
  }{7} Z^2\nonumber\\&+\frac{640}{7} W Y+\frac{648}{7} W Z+\frac{544}{7} Y Z \,.
\end{align}
The anomalous dimension reads
\begin{align}
\eta &=
\frac{1}{280}T^2+\frac{1}{70}T W+\frac{2}{35} T X+\frac{1}{35}T Y+\frac{1}{70}T Z+\frac{19
   }{1225}W^2+\frac{176}{1225} W X+\frac{76}{1225} W Y \nonumber\\&+\frac{1}{35}W Z+\frac{256
   }{245}X^2+\frac{128}{245} X Y+\frac{8}{49} X Z+\frac{108}{1225} Y^2+\frac{17}{245} Y
   Z+\frac{31}{1960} Z^2\,.
\end{align}

\subsubsection*{ \normalsize {\color{blue} {\rm{24-Cell}} $\{3, 4, 3\}$ }}

The $24$-cell is peculiar to the $N=4$ case. The independent polynomial invariants appear at order $k=2,6,8,12$ and they read
%
\begin{align}  
\rho_{\{3, 4, 3\}} &= 12 (\phi _1^2+\phi _2^2+\phi _3^2+\phi _4^2) \,,\label{rho24}\\
\tau_{\{3, 4, 3\}} &= 12 (\phi _1^6+5 \left(\phi _2^2+\phi _3^2+\phi _4^2\right) \phi _1^4+5
   \left(\phi _2^4+\phi _3^4+\phi _4^4\right) \phi _1^2\nonumber\\&+\phi _2^6+\phi
   _3^6+\phi _4^6+5 \phi _3^2 \phi _4^4+5 \phi _3^4 \phi _4^2+5 \phi _2^4
   \left(\phi _3^2+\phi _4^2\right)+5 \phi _2^2 \left(\phi _3^4+\phi
   _4^4\right)) \,,\label{tau24}\\
\sigma_{\{3, 4, 3\}} &= 4 (3 \phi _1^8+28 \left(\phi _2^2+\phi _3^2+\phi _4^2\right) \phi _1^6+70
   \left(\phi _2^4+\phi _3^4+\phi _4^4\right) \phi _1^4+28 \left(\phi
   _2^6+\phi _3^6+\phi _4^6\right) \phi _1^2\nonumber\\&+3 \phi _2^8+3 \phi _3^8+3 \phi
   _4^8+28 \phi _3^2 \phi _4^6+70 \phi _3^4 \phi _4^4+28 \phi _3^6 \phi
   _4^2+28 \phi _2^6 \left(\phi _3^2+\phi _4^2\right)\nonumber\\&+70 \phi _2^4 \left(\phi
   _3^4+\phi _4^4\right)+28 \phi _2^2 \left(\phi _3^6+\phi _4^6\right)) \,,\label{sigma24}\\
\omega_{\{3, 4, 3\}} &= 12 (\phi _1^{12}+22 \left(\phi _2^2+\phi _3^2+\phi _4^2\right) \phi
   _1^{10}+165 \left(\phi _2^4+\phi _3^4+\phi _4^4\right) \phi _1^8+308
   \left(\phi _2^6+\phi _3^6+\phi _4^6\right) \phi _1^6\nonumber\\&+165 \left(\phi
   _2^8+\phi _3^8+\phi _4^8\right) \phi _1^4+22 \left(\phi _2^{10}+\phi
   _3^{10}+\phi _4^{10}\right) \phi _1^2+\phi _2^{12}+\phi _3^{12}+\phi
   _4^{12}+22 \phi _3^2 \phi _4^{10}\nonumber\\&+165 \phi _3^4 \phi _4^8+308 \phi _3^6
   \phi _4^6+165 \phi _3^8 \phi _4^4+22 \phi _3^{10} \phi _4^2+22 \phi
   _2^{10} \left(\phi _3^2+\phi _4^2\right)+165 \phi _2^8 \left(\phi _3^4+\phi
   _4^4\right)\nonumber\\&+308 \phi _2^6 \left(\phi _3^6+\phi _4^6\right)+165 \phi _2^4
   \left(\phi _3^8+\phi _4^8\right)+22 \phi _2^2 \left(\phi _3^{10}+\phi
   _4^{10}\right))\,. \label{omega24}
\end{align}
%
It is natural to consider the critical behaviour of this system at the upper critical dimensions $d_c=3,\nicefrac{8}{3},\nicefrac{5}{2},\nicefrac{12}{5}$. Since they show a critical behaviour which is not ${\tt{O(N)}}$-like, we report the cases $d_c=\nicefrac{5}{2}$ and $d_c=\nicefrac{12}{5}$. 

We start considering the case $d_c=\nicefrac{5}{2}$. The marginal potential reads
 \begin{equation}
U(\rho,\tau,\sigma) = \frac{1}{10!} \left( X\, \rho^5_{\{3, 4, 3\}} + Y\, \rho^2_{\{3, 4, 3\}} \tau_{\{3, 4, 3\}} + Z\, \sigma_{\{3, 4, 3\}}\rho_{\{3, 4, 3\}}\right)\,,
\end{equation} 
and the corresponding LO beta functions are

\begin{align}
\beta_X &=-4\epsilon X + \frac{104398848}{35} X^2+\frac{173568}{7} X Y + 768 X Z +\frac{19876}{525} Y^2 + \frac{844}{225} Y Z + \frac{5642}{6075} Z^2 \,,\\
\beta_Y &=-4\epsilon Y  + \frac{108822528}{35}  X Y + \frac{2039808}{5} X Z + \frac{3320064}{175}  Y^2 + \frac{178816}{75} Y Z - \frac{121184}{675} Z^2 \,, \\
\beta_Z &= -4\epsilon Z + \frac{8552448}{5} X Z + \frac{4810752}{175} Y^2 + \frac{772096}{25} Y Z + \frac{930496}{225} Z^2\,,
\end{align}
with anomalous dimension
\begin{equation}
\eta=\frac{8153726976}{1225} X^2 + \frac{28311552}{245} X Y + \frac{2359296}{175}  X Z + \frac{3219456}{6125} Y^2 + \frac{16384}{125} Y Z + \frac{71168}{7875} Z^2\,.
\end{equation}
Beside the {\tt{penta-O(4)}} FP, there are two coincident 24-cell FPs characterised by the same anomalous dimension $\eta=0.0000115365$.

Finally we consider the theory in  $d_c = \nicefrac{12}{5}$. The marginal potential then is
\begin{equation}
U(\rho,\tau,\sigma,\omega) = \frac{1}{12!} \left( X\, \rho ^6 _{\{3, 4, 3\}} + Y\, \rho^3_{\{3, 4, 3\}} \tau_{\{3, 4, 3\}} + Z \rho^2_{\{3, 4, 3\}} \sigma_{\{3, 4, 3\}} + W\, \tau ^2_{\{3, 4, 3\}}+ T\, \omega_{\{3, 4, 3\}}    \right)\,,
\end{equation} 
and the LO beta functions read
%
\begin{align}  
 \beta_X &= -5\epsilon X
   -\frac{1925}{1492992} T^2 + \frac{17}{576}  T W + 560 T X+\frac{263}{54} T Y + \frac{1855}{3888} T Z - \frac{41567}{898128} W^2 \nonumber\\
   &+ \frac{43360}{231} W X - \frac{10313}{2079} W Y - \frac{4094}{8019} W Z + \frac{304349184}{11} X^2 + \frac{1420800}{7} X Y \nonumber\\
   &+ \frac{66560}{11}  X Z + \frac{45352}{231} Y^2 + \frac{292}{297} Y Z+\frac{25445}{16038} Z^2  \,,\\
\beta_Y &= -5\epsilon Y + \frac{385}{1296} T^2 - \frac{2398}{243} T W - 157440 T X - \frac{13076}{9}  T Y - \frac{36407}{243} T Z + \frac{537479}{37422} W^2 \nonumber\\
   &+\frac{30136320}{77} W X + \frac{998576}{297} W Y + \frac{155264}{891} W Z + \frac{2460450816}{77} X Y \nonumber\\
   &+ \frac{37232640}{11} X Z + \frac{14161536}{77} Y^2 + \frac{6724864}{297}  Y Z - \frac{599872}{2673} Z^2 \,,\\
\beta_Z  &=-5\epsilon Z -\frac{385}{288} T^2 + \frac{2809}{27} T W + 1658880 T X+14624 T Y+\frac{39452}{27} T Z + \frac{123734}{2079} W^2 \nonumber\\
&+\frac{4423680}{11} W X+\frac{2310208}{231}  W Y + \frac{92992}{33} W Z + 20348928 X Z+\frac{22076928}{77} Y^2 \nonumber\\
&+ \frac{9234944}{33} Y Z + \frac{7956928}{297} Z^2 \,,\\
\beta_W &=-5\epsilon W + \frac{77}{18} T^2 + \frac{2236}{27} T W + 3776 T Y+\frac{19712}{27} T Z+ \frac{86072}{297} W^2 \nonumber\\
&+ 2654208 W X + \frac{8694016}{231} W Y + \frac{559232}{99} W Z
   +\frac{43628544}{77} Y^2 \nonumber\\
   &+ \frac{7792640}{33} Y Z + \frac{7189504}{297} Z^2  \,,
      \end{align}
   \begin{align}
\beta_T &=-5\epsilon T+ \frac{2464}{3} T^2 + \frac{5536}{3} T W + 2654208 T X+92160 T Y+\frac{92288}{3} T Z +\frac{21760}{33} W^2 \nonumber \\
&+\frac{417792}{11} W Y + \frac{770048}{33} W Z + \frac{7766016}{11} Y Z + \frac{2265088}{11} Z^2 \,,
\end{align}
%
with anomalous dimension given by
\begin{equation}
\eta =
\frac{429981696}{245} X^2+\frac{1492992}{49} X Y+\frac{124416}{35} X Z+\frac{34992}{245} Y^2+\frac{1296}{35} Y Z+\frac{99}{35} Z^2\,.
\end{equation}
Apart from the {\tt Hexa-O(4)} fixed point the system displays two {\tt 24-Cell} fixed points whose critical exponents are given in Section \ref{UC}.

\subsubsection*{ \normalsize {\color{blue}  {\rm{600-Cell}} $\{5, 3, 3\}$ -  {\rm{120-Cell}} $\{3, 3, 5\}$ }}

The independent invariants have very complicated expressions and they appear at order $2,12,20,30$. The possible upper critical dimensions are therefore $d_c=\nicefrac{12}{5},\, \nicefrac{20}{9},\,\nicefrac{15}{7}$. We think it is not illuminating to report here the explicit expressions for the invariants as well as for the corresponding beta functions, but we point out that they can be extracted following the main lines of reasoning given in the main text.
%

\newpage
\section{Beta functionals}\label{Appendix:BetaFunctionals}

We report here the multicomponent beta functionals for the cases $d_c=\nicefrac{5}{2}$ and $d_c=\nicefrac{12}{5}$. In particular, in the $d_c=\nicefrac{12}{5}$ case we refer to $\gamma$ as the Euler's constant and to $\psi(z)=\Gamma'(z)/\Gamma(z)$ as the logarithmic derivative of the Gamma function.


\begin{figure}[h]
\begin{equation}\notag
d_c=\frac{5}{2}
\end{equation}
\begin{tikzpicture}
\draw (0,0) circle (.5cm);
\draw (-.5,0) to[out=50,in=130] (.5,0);
\draw (-.5,0) to[out=0,in=180] (.5,0);
\draw (-.5,0) to[out=-50,in=-130] (.5,0);
\filldraw [gray!50] (-.5,0) circle (2pt);
\draw (-.5,0) circle (2pt);
\filldraw [gray!50] (.5,0) circle (2pt);
\draw (.5,0) circle (2pt);
\draw (1.5,0) circle (.5cm);
\draw (1,0) to [out=50,in=130] (2,0);
\draw (1,0) to [out=75,in=180] (1.5,.375);
\draw (1,0) to [out=25,in=180] (1.5,.125);
\filldraw [gray!50] (1.5,.5) circle (2pt);
\draw (1.5,.5) circle (2pt);
\draw (1.5,.125) to [out=0,in=155] (2,0);
\draw (1,0) to [out=-25,in=180] (1.5,-.125);
\draw (1.5,-.125) to [out=0,in=-155] (2,0);
\draw (1.5,.375) to [out=0,in=115] (2,0);
\draw (1,0) to [out=-75,in=180] (1.5,-.375);
\draw (1.5,-.375) to [out=0,in=-115] (2,0);
\draw (1,0) to [out=0,in=180] (2,0);
\draw (1,0) to[out=-50,in=-130] (2,0);
\filldraw [gray!50] (1,0) circle (2pt);
\draw (1,0) circle (2pt);
\filldraw [gray!50] (2,0) circle (2pt);
\draw (2,0) circle (2pt);
\draw (3,0) circle (.5cm);
\draw (2.531,-.171) to [out=-30,in=-150] (3.469,-.171);
\draw (2.531,-.171) to[out=80,in=210] (3,.5);
\draw (2.531,-.171) to [out=35,in=250] (3,.5);
\draw (2.531,-.171) to [out=0,in=-90] (3,.3);
\draw (3,.3) to [out=90,in=-90] (3,.5);
\draw (3.469,-.171) to [out=180,in=-90] (3,.3);
\draw (3.469,-.171) to [out=100,in=-30] (3,.5);
\draw (3.469,-.171)  to [out=145,in=290] (3,.5);
\filldraw [gray!50] (3,.5) circle (2pt);
\draw (3,.5) circle (2pt);
\filldraw [gray!50] (3.469,-.171) circle (2pt);
\draw (3.469,-.171) circle (2pt);
\filldraw [gray!50] (2.531,-.171) circle (2pt);
\draw (2.531,-.171) circle (2pt);
\draw (4.5,0) circle (.5cm);
\draw (4.031,-.171) to [out=-45,in=180] (4.5,-.4);
\draw (4.5,-.4) to [out=0,in=225] (4.969,-.171);
\draw (4.031,-.171) to [out=80,in=-150] (4.5,.5);
\draw (4.969,-.171) to [out=100,in=-30] (4.5,.5);
\draw (4.031,-.171) to [out=-30,in=-150] (4.969,-.171);
\draw (4.031,-.171) to [out=30,in=150] (4.969,-.171);
\draw (4.031,-.171) to [out=45,in=180] (4.5,.075);
\draw (4.5,.075) to [out=0,in=135] (4.969,-.171);
\draw (4.031,-.171) to [out=0,in=180] (4.969,-.171);
\filldraw [gray!50] (4.5,.5) circle (2pt);
\draw (4.5,.5) circle (2pt);
\filldraw [gray!50] (4.969,-.171) circle (2pt);
\draw (4.969,-.171) circle (2pt);
\filldraw [gray!50] (4.031,-.171) circle (2pt);
\draw (4.031,-.171) circle (2pt);
\draw (6,0) circle (.5cm);
\draw (5.531,-.171) to [out=-30,in=-150] (6.469,-.171);
\draw (5.531,-.171) to [out=35,in=250] (6,.5);
\draw (5.531,-.171) to [out=80,in=210] (6,.5);
\draw (6.469,-.171)  to [out=100,in=-30] (6,.5);
\draw (6.469,-.171)  to [out=145,in=290] (6,.5);
\draw (5.531,-.171)  to [out=0,in=180] (6.469,-.171);
\draw (5.531,-.171)  to [out=30,in=150]  (6.469,-.171);
\filldraw [gray!50] (6,.5) circle (2pt);
\draw (6,.5) circle (2pt);
\filldraw [gray!50] (6.469,-.171) circle (2pt);
\draw (6.469,-.171) circle (2pt);
\filldraw [gray!50] (5.531,-.171) circle (2pt);
\draw (5.531,-.171) circle (2pt);
\draw (0,-1.25) circle (.5cm);
\draw (.469,-1.421) to [out=100,in=-30] (0,-.75);
\draw (-.469,-1.421) to [out=0,in=180] (.469,-1.421);
\draw (-.469,-1.421) to [out=65,in=180] (0,-1.08);
\draw (0,-1.08) to [out=0,in=115] (.469,-1.421);
\draw (-.469,-1.421) to [out=-45,in=180] (0,-1.65);
\draw (0,-1.65) to [out=0,in=225] (.469,-1.421);
\draw (-.469,-1.421) to [out=-30,in=-150] (.469,-1.421);
\draw (-.469,-1.421) to [out=30,in=150] (.469,-1.421);
\draw (-.469,-1.421) to [out=45,in=180] (0,-1.175);
\draw (0,-1.175) to [out=0,in=135] (.469,-1.421);
\filldraw [gray!50] (0,-.75) circle (2pt);
\draw (0,-.75) circle (2pt);
\filldraw [gray!50] (-.469,-1.421) circle (2pt);
\draw (-.469,-1.421) circle (2pt);
\filldraw [gray!50] (.469,-1.421) circle (2pt);
\draw (.469,-1.421) circle (2pt);
\draw (1.5,-1.25) circle (.5cm);
\draw (1.031,-1.421) to [out=-45,in=180] (1.5,-1.65);
\draw (1.5,-1.65) to [out=0,in=225] (1.969,-1.421);
\draw (1.969,-1.421) to [out=100,in=-30] (1.5,-.75);
\draw (1.969,-1.421) to [out=130,in=-60] (1.5,-.75);
\draw (1.031,-1.421) to [out=-30,in=-150] (1.969,-1.421);
\draw (1.031,-1.421) to [out=30,in=150] (1.969,-1.421);
\draw (1.031,-1.421) to [out=45,in=180] (1.5,-1.175);
\draw (1.5,-1.175) to [out=0,in=135] (1.969,-1.421);
\draw (1.031,-1.421) to [out=0,in=180] (1.969,-1.421);
\filldraw [gray!50] (1.5,-.75) circle (2pt);
\draw (1.5,-.75) circle (2pt);
\filldraw [gray!50] (1.031,-1.421) circle (2pt);
\draw (1.031,-1.421) circle (2pt);
\filldraw [gray!50] (1.969,-1.421) circle (2pt);
\draw (1.969,-1.421) circle (2pt);
\draw (3,-1.25) circle (.5cm);
\draw (2.531,-1.421) to [out=-45,in=180] (3,-1.65);
\draw (3,-1.65) to [out=0,in=225] (3.469,-1.421);
\draw (3.469,-1.421) to [out=100,in=-30] (3,-.75);
\draw (3.469,-1.421) to [out=130,in=-60] (3,-.75);
\draw (3.469,-1.421) to [out=150,in=-60] (3.1,-1.125);
\draw (3.1,-1.125) to [out=120,in=-90] (3,-.75);
\draw (2.531,-1.421) to [out=-30,in=-150] (3.469,-1.421);
\draw (2.531,-1.421) to [out=30,in=150] (3.469,-1.421);
\draw (2.531,-1.421) to [out=0,in=180] (3.469,-1.421);
\filldraw [gray!50] (3,-.75) circle (2pt);
\draw (3,-.75) circle (2pt);
\filldraw [gray!50] (2.531,-1.421) circle (2pt);
\draw (2.531,-1.421) circle (2pt);
\filldraw [gray!50] (3.469,-1.421) circle (2pt);
\draw (3.469,-1.421) circle (2pt);
\draw (4.5,-1.25) circle (.5cm);
\draw (4.031,-1.421) to [out=-45,in=180] (4.5,-1.65);
\draw (4.5,-1.65) to [out=0,in=225] (4.969,-1.421);
\draw (4.969,-1.421) to [out=100,in=-30] (4.5,-.75);
\draw (4.969,-1.421) to [out=130,in=-60] (4.5,-.75);
\draw (4.031,-1.421) to [out=80,in=210] (4.5,-.75);
\draw (4.031,-1.421) to [out=-30,in=-150] (4.969,-1.421);
\draw (4.031,-1.421) to [out=30,in=150] (4.969,-1.421);
\draw (4.031,-1.421) to [out=0,in=180] (4.969,-1.421);
\filldraw [gray!50] (4.5,-.75) circle (2pt);
\draw (4.5,-.75) circle (2pt);
\filldraw [gray!50] (4.031,-1.421) circle (2pt);
\draw (4.031,-1.421) circle (2pt);
\filldraw [gray!50] (4.969,-1.421) circle (2pt);
\draw (4.969,-1.421) circle (2pt);
\draw (6,-1.25) circle (.5cm);
\draw (5.5,-1.25) to [out=50,in=130] (6.5,-1.25);
\draw (5.5,-1.25) to [out=75,in=180] (6,-.85);
\draw (5.5,-1.25) to [out=25,in=180] (6,-1.125);
\draw (6,-1.125) to [out=0,in=155] (6.5,-1.25);
\draw (5.5,-1.25) to [out=-25,in=180] (6,-1.375);
\draw (6,-1.375) to [out=0,in=-155] (6.5,-1.25);
\draw (6,-.85) to [out=0,in=115] (6.5,-1.25);
\draw (5.5,-1.25) to [out=-75,in=180] (6,-1.65);
\draw (6,-1.65) to [out=0,in=-115] (6.5,-1.25);
\draw (5.5,-1.25) to [out=0,in=180] (6.5,-1.25);
\draw (5.5,-1.25) to[out=-50,in=-130] (6.5,-1.25);
\filldraw [blue!50] (5.5,-1.25) circle (2pt);
\draw (5.5,-1.25) circle (2pt);
\filldraw [blue!50] (6.5,-1.25) circle (2pt);
\draw (6.5,-1.25) circle (2pt);
\end{tikzpicture}
\begin{equation}\label{Betasd=5/2}
\begin{split}
\beta_{V}	&= \frac{1}{30}V_{a_1a_2a_3a_4a_5}V_{a_1a_2a_3a_4a_5} + \frac{1}{7560}V_{a_1a_2}V_{a_1a_3a_4a_5a_6a_7a_8a_9a_{10}}V_{a_2a_3a_4a_5a_6a_7a_8a_9a_{10}}\\
& -\frac{1}{144}\sqrt{\frac{\pi }{2}} \Gamma \left(\nicefrac{1}{4}\right)^2 V_{a_1a_2a_3a_4a_5a_6a_7a_8}V_{a_1a_2a_3a_4a_9a_{10}}V_{a_5a_6a_7a_8a_9a_{10}}\\
& +\frac{\Gamma \left(\nicefrac{3}{4}\right)^2}{45 \sqrt{2 \pi }}V_{a_1a_2a_3a_4}
   V_{a_1a_2a_5a_6a_7a_8a_9a_{10}}V_{a_3a_4a_5a_6a_7a_8a_9a_{10}}\\
   & -\frac{\pi  \Gamma \left(\nicefrac{1}{4}\right)^2}{216 \Gamma\left(\nicefrac{3}{4}\right)^2} V_{a_1a_2a_3a_4a_5a_6} V_{a_1a_2a_3a_7a_8a_9a_{10}} V_{a_4a_5a_6a_7a_8a_9a_{10}} \\
& +\frac{2}{945} V_{a_1a_2a_3} V_{a_1a_4a_5a_6a_7a_8a_9a_{10}}V_{a_2a_3a_4a_5a_6a_7a_8a_9a_{10}}\\
& +\frac{2}{135} V_{a_1a_2a_3a_4} V_{a_1a_5a_6a_7a_8a_9a_{10}} V_{a_2a_3a_4a_5a_6a_7a_8a_9a_{10}}\\
& -\frac{2}{45} V_{a_1a_2a_3a_4a_5}V_{a_1a_6a_7a_8a_9a_{10}} V_{a_2a_3a_4a_5a_6a_7a_8a_9a_{10}}\\
& +\frac{1}{45} \left[-4+\pi -\log (4)\right]  V_{a_1a_2a_3a_4a_5} V_{a_1a_2a_6a_7a_8a_9a_{10}}  V_{a_3a_4a_5a_6a_7a_8a_9a_{10}}\\ 
(\beta_{Z})_{a_1a_2}&= -\frac{1}{56700} V_{a_1a_3a_4a_5a_6a_7a_8a_9a_{10}a_{11}} V_{a_2a_3a_4a_5a_6a_7a_8a_9a_{10}a_{11}}
\end{split}
\end{equation}
\end{figure}
%


\begin{figure}
\begin{equation}\notag
d_c=\frac{12}{5}
\end{equation}
\begin{tikzpicture}
\draw (0,0) circle (.5cm);
\draw (-.5,0) to[out=30,in=150] (.5,0);
\draw (-.5,0) to[out=-30,in=-160] (.5,0);
\draw (-.5,0) to[out=60,in=180] (0,.32);
\draw (0,.32) to[out=0,in=130] (.5,0);
\draw (-.5,0) to[out=-60,in=180] (0,-.32);
\draw (0,-.32) to[out=0,in=-130] (.5,0);
\filldraw [gray!50] (-.5,0) circle (2pt);
\draw (-.5,0) circle (2pt);
\filldraw [gray!50] (.5,0) circle (2pt);
\draw (.5,0) circle (2pt);
\draw (1.5,0) circle (.5cm);
\draw (1,0) to [out=25,in=180] (1.5,.125);
\draw (1,0) to [out=50,in=130] (2,0);
\draw (1,0) to [out=62,in=180] (1.5,.3);
\draw (1,0) to [out=75,in=180] (1.5,.4);
\draw (1.5,.4) to [out=0,in=105] (2,0);
\draw (1.5,.3) to [out=0,in=118] (2,0);
\filldraw [gray!50] (1.5,.5) circle (2pt);
\draw (1.5,.5) circle (2pt);
\draw (1.5,.125) to [out=0,in=155] (2,0);
\draw (1,0) to [out=-25,in=180] (1.5,-.125);
\draw (1,0) to[out=-50,in=-130] (2,0);
\draw (1,0) to [out=-62,in=180] (1.5,-.3);
\draw (1,0) to [out=-75,in=180] (1.5,-.4);
\draw (1.5,-.125) to [out=0,in=-155] (2,0);
\draw (1.5,-.4) to [out=0,in=-115] (2,0);
\draw (1.5,-.3) to [out=0,in=-118] (2,0);
\draw (1,0) to [out=0,in=180] (2,0);
\filldraw [gray!50] (1,0) circle (2pt);
\draw (1,0) circle (2pt);
\filldraw [gray!50] (2,0) circle (2pt);
\draw (2,0) circle (2pt);
\draw (3,0) circle (.5cm);
\draw (2.531,-.171) to [out=0,in=-90] (3,.3);
\draw (2.531,-.171) to [out=30,in=250] (3,.5);
\draw (2.531,-.171) to[out=55,in=230] (3,.5);
\draw (2.531,-.171) to[out=80,in=210] (3,.5);
\draw (3,.3) to [out=90,in=-90] (3,.5);
\draw (3.469,-.171) to [out=180,in=-90] (3,.3);
\draw (3.469,-.171)  to [out=150,in=290] (3,.5);
\draw (3.469,-.171) to [out=125,in=-50] (3,.5);
\draw (3.469,-.171) to [out=100,in=-30] (3,.5);
\draw (2.531,-.171) to [out=-30,in=-150] (3.469,-.171);
\filldraw [gray!50] (3,.5) circle (2pt);
\draw (3,.5) circle (2pt);
\filldraw [gray!50] (3.469,-.171) circle (2pt);
\draw (3.469,-.171) circle (2pt);
\filldraw [gray!50] (2.531,-.171) circle (2pt);
\draw (2.531,-.171) circle (2pt);
\draw (4.5,0) circle (.5cm);
\draw (4.031,-.171) to [out=0,in=180] (4.969,-.171);
\draw (4.031,-.171) to [out=-45,in=180] (4.5,-.425);
\draw (4.5,-.425) to [out=0,in=225] (4.969,-.171);
\draw (4.031,-.171) to [out=-30,in=180] (4.5,-.325);
\draw (4.5,-.325) to [out=0,in=210] (4.969,-.171);
\draw (4.031,-.171) to [out=-15,in=180] (4.5,-.25);
\draw (4.5,-.25) to [out=0,in=195] (4.969,-.171);
\draw (4.031,-.171) to [out=45,in=180] (4.5,.083);
\draw (4.5,.083) to [out=0,in=135] (4.969,-.171);
\draw (4.031,-.171) to [out=30,in=180] (4.5,-.017);
\draw (4.5,-.017) to [out=0,in=150] (4.969,-.171);
\draw (4.031,-.171) to [out=15,in=180] (4.5,.-.092);
\draw (4.5,.-.092) to [out=0,in=165] (4.969,-.171);
\draw (4.031,-.171) to [out=80,in=-150] (4.5,.5);
\draw (4.969,-.171) to [out=100,in=-30] (4.5,.5);
\filldraw [gray!50] (4.5,.5) circle (2pt);
\draw (4.5,.5) circle (2pt);
\filldraw [gray!50] (4.969,-.171) circle (2pt);
\draw (4.969,-.171) circle (2pt);
\filldraw [gray!50] (4.031,-.171) circle (2pt);
\draw (4.031,-.171) circle (2pt);
\draw (6,0) circle (.5cm);
\draw (5.531,-.171) to [out=35,in=250] (6,.5);
\draw (5.531,-.171) to [out=80,in=210] (6,.5);
\draw (6.469,-.171)  to [out=100,in=-30] (6,.5);
\draw (6.469,-.171)  to [out=145,in=290] (6,.5);
\draw (5.531,-.171) to [out=-30,in=180] (6,-.325);
\draw (6,-.325) to [out=0,in=210] (6.469,-.171);
\draw (5.531,-.171) to [out=-15,in=180] (6,-.25);
\draw (6,-.25) to [out=0,in=195] (6.469,-.171);
\draw (5.531,-.171) to [out=30,in=180] (6,-.017);
\draw (6,-.017) to [out=0,in=150] (6.469,-.171);
\draw (5.531,-.171) to [out=15,in=180] (6,.-.092);
\draw (6,.-.092) to [out=0,in=165] (6.469,-.171);
\draw (5.531,-.171)  to [out=0,in=180] (6.469,-.171);
\filldraw [gray!50] (6,.5) circle (2pt);
\draw (6,.5) circle (2pt);
\filldraw [gray!50] (6.469,-.171) circle (2pt);
\draw (6.469,-.171) circle (2pt);
\filldraw [gray!50] (5.531,-.171) circle (2pt);
\draw (5.531,-.171) circle (2pt);
\draw (7.5,0) circle (.5cm);
\draw (7.969,-.171) to [out=100,in=-30] (7.5,.5);
\draw (7.031,-.171) to [out=65,in=180] (7.5,0.17);
\draw (7.5,0.17) to [out=0,in=115] (7.969,-.171);
\draw (7.031,-.171) to [out=0,in=180] (7.969,-.171);
\draw (7.031,-.171) to [out=-45,in=180] (7.5,-.425);
\draw (7.5,-.425) to [out=0,in=225] (7.969,-.171);
\draw (7.031,-.171) to [out=-30,in=180] (7.5,-.325);
\draw (7.5,-.325) to [out=0,in=210] (7.969,-.171);
\draw (7.031,-.171) to [out=-15,in=180] (7.5,-.25);
\draw (7.5,-.25) to [out=0,in=195] (7.969,-.171);
\draw (7.031,-.171) to [out=45,in=180] (7.5,.083);
\draw (7.5,.083) to [out=0,in=135] (7.969,-.171);
\draw (7.031,-.171) to [out=30,in=180] (7.5,-.017);
\draw (7.5,-.017) to [out=0,in=150] (7.969,-.171);
\draw (7.031,-.171) to [out=15,in=180] (7.5,.-.092);
\draw (7.5,.-.092) to [out=0,in=165] (7.969,-.171);
\filldraw [gray!50] (7.5,.5) circle (2pt);
\draw (7.5,.5) circle (2pt);
\filldraw [gray!50] (7.969,-.171) circle (2pt);
\draw (7.969,-.171) circle (2pt);
\filldraw [gray!50] (7.031,-.171) circle (2pt);
\draw (7.031,-.171) circle (2pt);
\draw (9,0) circle (.5cm);
\draw (9.469,-.171) to [out=100,in=-30] (9,.5);
\draw (9.469,-.171) to [out=130,in=-60] (9,.5);
\draw (8.531,-.171) to [out=0,in=180] (9.469,-.171);
\draw (8.531,-.171) to [out=-45,in=180] (9,-.425);
\draw (9,-.425) to [out=0,in=225] (9.469,-.171);
\draw (8.531,-.171) to [out=-30,in=180] (9,-.325);
\draw (9,-.325) to [out=0,in=210] (9.469,-.171);
\draw (8.531,-.171) to [out=-15,in=180] (9,-.25);
\draw (9,-.25) to [out=0,in=195] (9.469,-.171);
\draw (8.531,-.171) to [out=45,in=180] (9,.083);
\draw (9,.083) to [out=0,in=135] (9.469,-.171);
\draw (8.531,-.171) to [out=30,in=180] (9,-.017);
\draw (9,-.017) to [out=0,in=150] (9.469,-.171);
\draw (8.531,-.171) to [out=15,in=180] (9,.-.092);
\draw (9,.-.092) to [out=0,in=165] (9.469,-.171);
\filldraw [gray!50] (9,.5) circle (2pt);
\draw (9,.5) circle (2pt);
\filldraw [gray!50] (8.531,-.171) circle (2pt);
\draw (8.531,-.171) circle (2pt);
\filldraw [gray!50] (9.469,-.171) circle (2pt);
\draw (9.469,-.171) circle (2pt);
\draw (0,-1.25) circle (.5cm);
\draw (-.469,-1.421) to [out=0,in=180] (.469,-1.421);
\draw (-.469,-1.421) to [out=-45,in=180] (0,-1.675);
\draw (0,-1.675) to [out=0,in=225] (.469,-1.421);
\draw (-.469,-1.421) to [out=-30,in=180] (0,-1.575);
\draw (0,-1.575) to [out=0,in=210] (.469,-1.421);
\draw (-.469,-1.421) to [out=-15,in=180] (0,-1.5);
\draw (0,-1.5) to [out=0,in=195] (.469,-1.421);
\draw (-.469,-1.421) to [out=30,in=180] (0,-1.267);
\draw (0,-1.267) to [out=0,in=150] (.469,-1.421);
\draw (-.469,-1.421) to [out=15,in=180] (0,-1.342);
\draw (0,-1.342) to [out=0,in=165] (.469,-1.421);
\draw (.469,-1.421) to [out=150,in=290] (0,-.75);
\draw (.469,-1.421) to [out=125,in=-50] (0,-.75);
\draw (.469,-1.421) to [out=100,in=-30] (0,-.75);
\filldraw [gray!50] (0,-.75) circle (2pt);
\draw (0,-.75) circle (2pt);
\filldraw [gray!50] (-.469,-1.421) circle (2pt);
\draw (-.469,-1.421) circle (2pt);
\filldraw [gray!50] (.469,-1.421) circle (2pt);
\draw (.469,-1.421) circle (2pt);
\draw (1.5,-1.25) circle (.5cm);
\draw (1.031,-1.421) to [out=30,in=250] (1.5,-.75);
\draw (1.031,-1.421) to[out=55,in=230] (1.5,-.75);
\draw (1.031,-1.421) to[out=80,in=210] (1.5,-.75);
\draw (1.969,-1.421)  to [out=150,in=290] (1.5,-.75);
\draw (1.969,-1.421) to [out=125,in=-50] (1.5,-.75);
\draw (1.969,-1.421) to [out=100,in=-30] (1.5,-.75);
\draw (1.031,-1.421) to [out=-20,in=-160] (1.969,-1.421);
\draw (1.031,-1.421) to [out=0,in=180] (1.969,-1.421);
\draw (1.031,-1.421) to [out=20,in=160] (1.969,-1.421);
\filldraw [gray!50] (1.5,-.75) circle (2pt);
\draw (1.5,-.75) circle (2pt);
\filldraw [gray!50] (1.031,-1.421) circle (2pt);
\draw (1.031,-1.421) circle (2pt);
\filldraw [gray!50] (1.969,-1.421) circle (2pt);
\draw (1.969,-1.421) circle (2pt);
\draw (3,-1.25) circle (.5cm);
\draw (2.531,-1.421) to [out=0,in=180] (3.469,-1.421);
\draw (2.531,-1.421)  to [out=-45,in=180] (3,-1.675);
\draw (3,-1.675) to [out=0,in=225] (3.469,-1.421);
\draw (2.531,-1.421)  to [out=-30,in=180] (3,-1.575);
\draw (3,-1.575) to [out=0,in=210] (3.469,-1.421);
\draw (2.531,-1.421)  to [out=-15,in=180] (3,-1.5);
\draw (3,-1.5) to [out=0,in=195] (3.469,-1.421);
\draw (2.531,-1.421)  to [out=15,in=180] (3,-1.342);
\draw (3,-1.342) to [out=0,in=165] (3.469,-1.421);
\draw (3.05,-1.1) to [out=110,in=-90] (3,-.75);
\draw (3.469,-1.421) to [out=160,in=-70] (3.05,-1.1);
\draw (3.469,-1.421)  to [out=150,in=290] (3,-.75);
\draw (3.469,-1.421) to [out=125,in=-50] (3,-.75);
\draw (3.469,-1.421) to [out=100,in=-30] (3,-.75);
\filldraw [gray!50] (3,-.75) circle (2pt);
\draw (3,-.75) circle (2pt);
\filldraw [gray!50] (2.531,-1.421) circle (2pt);
\draw (2.531,-1.421) circle (2pt);
\filldraw [gray!50] (3.469,-1.421) circle (2pt);
\draw (3.469,-1.421) circle (2pt);
\draw (4.5,-1.25) circle (.5cm);
\draw (4.031,-1.421) to [out=0,in=180] (4.969,-1.421);
\draw (4.031,-1.421) to [out=-45,in=180] (4.5,-1.675);
\draw (4.5,-1.675) to [out=0,in=225] (4.969,-1.421);
\draw (4.031,-1.421) to [out=-30,in=180] (4.5,-1.575);
\draw (4.5,-1.575) to [out=0,in=210] (4.969,-1.421);
\draw (4.031,-1.421) to [out=-15,in=180] (4.5,-1.5);
\draw (4.5,-1.5) to [out=0,in=195] (4.969,-1.421);
\draw (4.031,-1.421) to [out=30,in=180] (4.5,-1.267);
\draw (4.5,-1.267) to [out=0,in=150] (4.969,-1.421);
\draw (4.031,-1.421) to [out=15,in=180] (4.5,.-1.342);
\draw (4.5,.-1.342) to [out=0,in=165] (4.969,-1.421);
\draw (4.031,-1.421) to [out=80,in=-150] (4.5,-.75);
\draw (4.969,-1.421) to [out=100,in=-30] (4.5,-.75);
\draw (4.969,-1.421)  to [out=150,in=290] (4.5,-.75);
\filldraw [gray!50] (4.5,-.75) circle (2pt);
\draw (4.5,-.75) circle (2pt);
\filldraw [gray!50] (4.969,-1.421) circle (2pt);
\draw (4.969,-1.421) circle (2pt);
\filldraw [gray!50] (4.031,-1.421) circle (2pt);
\draw (4.031,-1.421) circle (2pt);
\draw (6,-1.25) circle (.5cm);
\draw (5.531,-1.421) to [out=80,in=210] (6,-.75);
\draw (6.469,-1.421)  to [out=100,in=-30] (6,-.75);
\draw (6.469,-1.421)  to [out=145,in=290] (6,-.75);
\draw (6.469,-1.421) to [out=125,in=-50] (6,-.75);
\draw (5.531,-1.421) to [out=0,in=180] (6.469,-1.421);
\draw (5.531,-1.421) to [out=-30,in=180] (6,-1.575);
\draw (6,-1.575) to [out=0,in=210] (6.469,-1.421);
\draw (5.531,-1.421) to [out=-15,in=180] (6,-1.5);
\draw (6,-1.5) to [out=0,in=195] (6.469,-1.421);
\draw (5.531,-1.421) to [out=30,in=180] (6,-1.267);
\draw (6,-1.267) to [out=0,in=150] (6.469,-1.421);
\draw (5.531,-1.421) to [out=15,in=180] (6,.-1.342);
\draw (6,.-1.342) to [out=0,in=165] (6.469,-1.421);
\filldraw [gray!50] (6,-.75) circle (2pt);
\draw (6,-.75) circle (2pt);
\filldraw [gray!50] (6.469,-1.421) circle (2pt);
\draw (6.469,-1.421) circle (2pt);
\filldraw [gray!50] (5.531,-1.421) circle (2pt);
\draw (5.531,-1.421) circle (2pt);
\draw (7.5,-1.25) circle (.5cm);
\draw (7.031,-1.421) to [out=80,in=210] (7.5,-.75);
\draw (7.969,-1.421)  to [out=100,in=-30] (7.5,-.75);
\draw (7.969,-1.421)  to [out=145,in=290] (7.5,-.75);
\draw (7.969,-1.421) to [out=125,in=-50] (7.5,-.75);
\draw (7.031,-1.421) to [out=-30,in=180] (7.5,-1.575);
\draw (7.5,-1.575) to [out=0,in=210] (7.969,-1.421);
\draw (7.031,-1.421) to [out=-15,in=180] (7.5,-1.5);
\draw (7.5,-1.5) to [out=0,in=195] (7.969,-1.421);
\draw (7.031,-1.421) to [out=30,in=180] (7.5,-1.267);
\draw (7.5,-1.267) to [out=0,in=150] (7.969,-1.421);
\draw (7.031,-1.421) to [out=15,in=180] (7.5,.-1.342);
\draw (7.5,.-1.342) to [out=0,in=165] (7.969,-1.421);
\draw (7.031,-1.421) to [out=30,in=250] (7.5,-.75);
\filldraw [gray!50] (7.5,-.75) circle (2pt);
\draw (7.5,-.75) circle (2pt);
\filldraw [gray!50] (7.969,-1.421) circle (2pt);
\draw (7.969,-1.421) circle (2pt);
\filldraw [gray!50] (7.031,-1.421) circle (2pt);
\draw (7.031,-1.421) circle (2pt);
\draw (9,-1.25) circle (.5cm);
\draw (8.5,-1.25) to [out=25,in=180] (9,-1.125);
\draw (8.5,-1.25) to [out=50,in=130] (9.5,-1.25);
\draw (8.5,-1.25) to [out=62,in=180] (9,-.95);
\draw (8.5,-1.25) to [out=75,in=180] (9,-.85);
\draw (9,-.85) to [out=0,in=105] (9.5,-1.25);
\draw (9,-.95) to [out=0,in=118] (9.5,-1.25);
\draw (9,-1.125) to [out=0,in=155] (9.5,-1.25);
\draw (8.5,-1.25) to [out=-25,in=180] (9,-1.375);
\draw (8.5,-1.25) to[out=-50,in=-130] (9.5,-1.25);
\draw (8.5,-1.25) to [out=-62,in=180] (9,-1.55);
\draw (8.5,-1.25) to [out=-75,in=180] (9,-1.65);
\draw (9,-1.3755) to [out=0,in=-155] (9.5,-1.25);
\draw (9,-1.65) to [out=0,in=-115] (9.5,-1.25);
\draw (9,-1.55) to [out=0,in=-118] (9.5,-1.25);
\draw (8.5,-1.25) to [out=0,in=180] (9.5,-1.25);
\filldraw [blue!50] (8.5,-1.25) circle (2pt);
\draw (8.5,-1.25) circle (2pt);
\filldraw [blue!50] (9.5,-1.25) circle (2pt);
\draw (9.5,-1.25) circle (2pt);
\end{tikzpicture}
\begin{equation}\label{Betasd=12/5}
\begin{split}
\beta_{V}	&=\frac{1}{144}V_{a_1a_2a_3a_4a_5a_6}V_{a_1a_2a_3a_4a_5a_6}+\frac{1}{580608}V_{a_1a_2}V_{a_1a_3a_4a_5a_6a_7a_8a_9a_{10}a_{11}a_{12}}V_{a_2a_3a_4a_5a_6a_7a_8a_9a_{10}a_{11}a_{12}}\\
   &-\frac{\Gamma \left(\nicefrac{1}{5}\right)^3
   \Gamma \left(\nicefrac{4}{5}\right) }{5760 \Gamma
   \left(\frac{2}{5}\right)} V_{a_1a_2a_3a_4a_5a_6a_7a_8a_9a_{10}}V_{a_1a_2a_3a_4a_5a_{11}a_{12}}V_{a_6a_7a_8a_9a_{10}a_{11}a_{12}}\\
   &-\frac{\Gamma \left(-\nicefrac{2}{5}\right) \Gamma
   \left(\nicefrac{1}{5}\right) \Gamma \left(\nicefrac{4}{5}\right)^2 }{32256 \Gamma \left(\nicefrac{2}{5}\right)^2 \Gamma
   \left(\nicefrac{8}{5}\right)}V_{a_1a_2a_3a_4}V_{a_1a_2a_5a_6a_7a_8a_9a_{10}a_{11}a_{12}}V_{a_3a_4a_5a_6a_7a_8a_9a_{10}a_{11}a_{12}}\\
   &+\frac{5 \left[-5+\gamma -\psi\left(\nicefrac{1}{5}\right)+2
   \psi\left(\nicefrac{3}{5}\right)\right]}{5184}V_{a_1a_2a_3a_4a_5a_6} V_{a_1a_2a_3a_7a_8a_9a_{10}a_{11}a_{12}}V_{a_4a_5a_6a_7a_8a_9a_{10}a_{11}a_{12}}\\
   &+\frac{25}{870912}V_{a_1a_2a_3} V_{a_1a_4a_5a_6a_7a_8a_9a_{10}a_{11}a_{12}}V_{a_2a_3a_4a_5a_6a_7a_8a_9a_{10}a_{11}a_{12}}\\
   &+\frac{25}{145152}V_{a_1a_2a_3a_4} V_{a_1a_5a_6a_7a_8a_9a_{10}a_{11}a_{12}}V_{a_2a_3a_4a_5a_6a_7a_8a_9a_{10}a_{11}a_{12}}\\
   &+\frac{25}{24192} V_{a_1a_2a_3a_4a_5} V_{a_1a_6a_7a_8a_9a_{10}a_{11}a_{12}}V_{a_2a_3a_4a_5a_6a_7a_8a_9a_{10}a_{11}a_{12}}\\
   &-\frac{5 \Gamma \left(\nicefrac{1}{5}\right) \Gamma \left(\nicefrac{2}{5}\right)^3
   }{41472 \Gamma \left(\nicefrac{4}{5}\right)^3}V_{a_1a_2a_3a_4a_5a_6a_7a_8}V_{a_1a_2a_3a_4a_9a_{10}a_{11}a_{12}}V_{a_5a_6a_7a_8a_9a_{10}a_{11}a_{12}}\\
   &-\frac{5}{1728}V_{a_1a_2a_3a_4a_5a_6} V_{a_1a_7a_8a_9a_{10}a_{11}a_{12}}V_{a_2a_3a_4a_5a_6a_7a_8a_9a_{10}a_{11}a_{12}}\\
   &-\frac{\Gamma \left(-\nicefrac{1}{5}\right) \Gamma
   \left(\nicefrac{1}{5}\right) \Gamma \left(\nicefrac{4}{5}\right)}{6048 \Gamma \left(\nicefrac{2}{5}\right) \Gamma
   \left(\nicefrac{7}{5}\right)}V_{a_1a_2a_3a_4a_5} V_{a_1a_2a_6a_7a_8a_9a_{10}a_{11}a_{12}}V_{a_3a_4a_5a_6a_7a_8a_9a_{10}a_{11}a_{12}}\\
   &+\frac{5 \left[-5+\gamma -\psi\left(\nicefrac{1}{5}\right)+\psi\left(\nicefrac{2}{5}\right)+\psi\left(\nicefrac{4}{5}\right)\right]}{3456}V_{a_1a_2a_3a_4a_5a_6} V_{a_1a_2a_7a_8a_9a_{10}a_{11}a_{12}}V_{a_3a_4a_5a_6a_7a_8a_9a_{10}a_{11}a_{12}}\\
   &-\frac{\Gamma \!\left(\nicefrac{1}{5}\right)^2 \Gamma \!\left(\nicefrac{2}{5}\right)}{1728 \,\Gamma \!\left(\nicefrac{4}{5}\right)} V_{a_1a_2a_3a_4a_5a_6a_7} V_{a_1a_2a_3a_8a_9a_{10}a_{11}a_{12}}V_{a_4a_5a_6a_7a_8a_9a_{10}a_{11}a_{12}}\\ 
(\beta_{Z})_{a_1a_2}&= -\frac{1}{4790016} V_{a_1a_3a_4a_5a_6a_7a_8a_9a_{10}a_{11}a_{12}a_{13}} V_{a_2a_3a_4a_5a_6a_7a_8a_9a_{10}a_{11}a_{12}a_{13}}
\end{split}
\end{equation}
\end{figure}

\clearpage
\enlargethispage{2\baselineskip}

\bibliography{PlatonicBib}

\end{document}